\documentclass[12pt]{article}

\usepackage[normalem]{ulem}
\usepackage[mathscr]{eucal}
\usepackage[version=3]{mhchem}

\usepackage{epsfig,latexsym,amsfonts,amsmath,amsthm,amssymb,amsbsy,multirow,slashed,wasysym,textcomp,dsfont,comment,mathtools,cancel,cite,datetime,appendix,BOONDOX-calo}
\usepackage{tocloft}
\usepackage{bbold}
\usepackage{tikz}
\usetikzlibrary{backgrounds}
\usetikzlibrary{patterns}
\usetikzlibrary{arrows.meta}
\tikzstyle{bag} = [align=center]
\usetikzlibrary{decorations.pathmorphing}
\def\bea{\begin{eqnarray}}
\def\eea{\end{eqnarray}}

\usepackage{hyperref}
\usepackage{subcaption}

 \newcommand{\badat}{\begin{alignedat}}
 \newcommand{\eadat}{\end{alignedat}}
 \newcommand\scalemath[2]{\scalebox{#1}{\mbox{\ensuremath{\displaystyle #2}}}}
 \def\be{\begin{equation}}
\def\ee{\end{equation}}

\def\p{\partial}

\usepackage{xcolor}

\newcommand{\pink}[1]{\textcolor{\pink}{#1}}

\definecolor{dblue}{rgb}{0.2,0.50,0.80}

\def\O{\mathcal{O}}

\def\bz{{\bar z}}
\def\bw{{\bar w}}

\def\ba{{\bar a}}
\def\bb{{\bar b}}
\def\bc{{\bar c}}
\def\bd{{\bar d}}

\def\bz{{\bar z}}
\def\bw{{\bar w}}

\def\pa{{\partial}}
\def\a{{\alpha}}

\DeclareFontFamily{OT1}{pzc}{}
\DeclareFontShape{OT1}{pzc}{m}{it}{<-> s * [1.10] pzcmi7t}{}
\DeclareMathAlphabet{\mathpzc}{OT1}{pzc}{m}{it}

\definecolor{vert}{rgb}{0.1367 0.543 0.1367}

\numberwithin{equation}{section} 

\begin{document}

 \begin{titlepage}
  \thispagestyle{empty}
  \begin{flushright}
  \end{flushright}
  \bigskip
  \begin{center}

        \baselineskip=13pt {\LARGE \scshape{
       Celestial Conformal Colliders}}
     
      \vskip1cm 

   \centerline{ 
   {Yangrui Hu} 
    and {Sabrina Pasterski}
}

\bigskip\bigskip
 
 \centerline{
\it Perimeter Institute for Theoretical Physics, Waterloo, ON N2L 2Y5, Canada}

\bigskip\bigskip

\end{center}

\begin{abstract}
 \noindent 
 We start by observing that the light-ray operators featured in the conformal collider literature are celestial primaries.
 This allows us to rephrase the corresponding 4D CFT correlators as probing a conformally soft matter sector of the 2D celestial CFT (CCFT). To demonstrate the utility of this perspective we show how the recent $w_{1+\infty}$ symmetry observed in CCFT suggests a natural extension of the conformal collider operators.

\end{abstract}

\end{titlepage}

\tableofcontents

\section{Introduction}

Celestial holography aims to apply the holographic principle to asymptotically flat spacetimes, encoding the gravitational $\cal S$-matrix in terms of a CFT living on the celestial sphere.
A main source of excitement in this 
program~\cite{Pasterski:2021raf,Pasterski:2021rjz,Raclariu:2021zjz} is that we're approaching a beautiful merger of techniques from the relativity, bootstrap, and amplitudes communities. The aim of this paper is to hone in on a particular application that highlights how insights from celestial CFT (CCFT) can inform discussions in these ventures.

Recent progress towards constructing a flat space hologram has centered around symmetry enhancements in the infrared. These come in two forms. First, the asymptotic symmetry group of asymptotically flat spacetimes promotes the standard global charges to an angular dependent form, at the expense of conservation laws now taking the form of flux balance laws~\cite{Bondi:1962px,Sachs:1962wk,Sachs:1962zza}. An important insight of Strominger was that the Ward identities for these symmetries could be reinterpreted as well-known soft theorems in QFT~\cite{Strominger:2017zoo}. A natural question that then arises is: how many soft theorems are there? This leads to a second infinity of symmetries. By going to a basis of boost eigenstates~\cite{Pasterski:2016qvg,Pasterski:2017kqt,Pasterski:2017ylz}, it appears that there exists an entire tower of soft theorems corresponding to poles in the complex conformal weight plane~\cite{Guevara:2019ypd}. 

While the `soft physics' investigations have naturally focused on (conformally) soft gauge bosons and gravitons, their collinear limits, and symmetry interpretations, the matter modes that they couple to are equally important. Indeed, in~\cite{Cordova:2018ygx} Cordova and Shao nicely demonstrated that they could realize the BMS algebra via light-ray operators constructed from the stress tensor on a fixed light sheet in any unitary (bulk) CFT. This has a natural application whenever the bulk is perturbative, and we consider such operators living on null infinity. While this predates most of the literature on conformally soft theorems~\cite{Pate:2019mfs,Puhm:2019zbl,Adamo:2019ipt}, with \cite{Cheung:2016iub} being a notable exception, in hindsight we can recognize these matter modes as being weight $\Delta=s+1$ primaries that source generalized currents constructed from the spin-$s$ gauge modes in CCFT (see the `celestial diamonds' of~\cite{Pasterski:2021fjn,Pasterski:2021dqe} building off observations in~\cite{Banerjee:2018fgd,Banerjee:2019aoy,Banerjee:2019tam}). Namely, the 4D light-ray operators are the `hard charges' appearing in the Ward identities we get when we couple the theory to gravity.

Importantly, each of these light-ray-supported matter symmetry generators has a definite boost weight! Despite this being mentioned in passing in the seminal work of Hofman and Maldacena~\cite{Hofman:2008ar}, there remains a disconnect between the very rich CFT literature building of this work and the celestial efforts. A primary goal of this paper is to fill in this gap. Upon observing that the light-ray operators featured in the conformal collider literature are celestial primaries, we can rephrase the corresponding 4D CFT correlators as probing a conformally soft matter sector of the 2D celestial CFT. We are then in a position to ask if and how recent CCFT results can inform the conformal collider discussion. Here we show that the recent $w_{1+\infty}$ symmetry observed in CCFT~\cite{Guevara:2021abz,Strominger:2021mtt,Adamo:2021lrv} suggests a natural extension of the conformal collider operators.

This paper is organized as follows. We begin section~\ref{sec:LRcorner} by reviewing salient aspects of the conformal collider literature, before showing that the light-ray 
operators featured there are also celestial primaries. Section~\ref{sec:tower} then focuses on how the celestial story points to further symmetry enhancements, which in turn point to corresponding towers of detector/light-ray operators in the matter sector. We close with a discussion of how the various programs we encounter intertwine in section~\ref{sec:discussion} and include additional computational details in the appendix.

\section{Light-ray Operators at the Corner}\label{sec:LRcorner}

Conformal collider physics applies CFT techniques and insights from AdS/CFT to compute observables relevant to particle experiments. This program was initiated by Hofman and Maldacena in 2008~\cite{Hofman:2008ar}, where they considered correlation functions of  average null energy (ANEC) operators ${\cal E}(\vec{n})$ at null infinity. To a good approximation, we can think of one of these operators sitting at a point on the celestial sphere $\vec{n}\in S^{d-2}$ as measuring the energy deposited in a physical calorimeter surrounding a particle collision in that fixed direction. For $d=4$ these take the form
\begin{equation}
\begin{split}
{\cal E}(\vec{n}) ~=&~ \lim\limits_{r\to\infty}\,r^{2}\,\int_{-\infty}^{\infty}\,du\,T_{uu}(u,r,\vec{n})~,
\end{split}
\label{equ:ANE-def-x}
\end{equation}
where the $r^2$ prefactor in this definition is to have a well-defined limit as $r\to\infty$. Similarly, a charge flux operator can be defined by replacing the stress tensor with the global $U(1)$ symmetry current. Notably, hadronic energy patterns produced in the electron-positron annihilation $e^+e^-\to N$ hadrons were used as early precision tests of QCD in the 1970s~\cite{Basham:1978bw,Basham:1978zq}.

Causality, a la the eponymous Average Null Energy Condition, gives us a positivity condition for expectation values of these operators
\be
\langle \Psi|{\cal E}|\Psi\rangle\ge0
\ee
that places constraints on field theories~\cite{Camanho:2009vw,Hoyos:2010at,Li:2015itl,Hartman:2016lgu,Hofman:2016awc,Faulkner:2016mzt,Balakrishnan:2017bjg,Cordova:2017zej,Kravchuk:2018htv,Ceyhan:2018zfg}. Many interesting directions have been studied extensively in the conformal collider literature, including energy correlators and event shapes~\cite{Belitsky:2013ofa,Belitsky:2013bja,Belitsky:2013xxa,Dixon:2019uzg,Chen:2019bpb,Gonzo:2020xza,Chang:2022ryc,Lee:2022ige}, light-ray OPEs and celestial blocks\footnote{Not to be conflated with studies of conformal block expansions in celestial CFT~\cite{Atanasov2021cje,Guevara:2021tvr,Hu:2022syq,De:2022gjn}}~\cite{Kologlu:2019mfz,Chen:2022jhb,Chang:2022ryc,Chang:2020qpj}, light-ray operator algebras~\cite{Casini:2017roe,Cordova:2018ygx,Gonzo:2020xza,Belin:2020lsr}, and detectors~\cite{Gonzo:2020xza,Caron-Huot:2022eqs}. For many of these applications it is convenient to rephrase the ANEC operator as a light transform of the stress tensor~\cite{Kravchuk:2018htv,Kologlu:2019mfz,Chang:2020qpj}
\begin{equation}
    {\cal E}(\vec{n})  ~=~ 2\, {\rm \bf L}[T(X_{i^0},Z_{\infty})] ~,
\end{equation}
where, in terms of the embedding space position $X$ and polarization $Z$\footnote{We refer interested readers to~\cite{Kravchuk:2018htv,Kologlu:2019mfz,Chang:2020qpj} for background on (and notation used in) the embedding space formalism as well as detailed discussions of light transforms and light-ray operators in Lorentzian CFTs.} we have
\begin{equation}
    {\rm \bf L}[{\cal O}(X,Z)] ~=~ \int_{-\infty}^{\infty}\,d\a \,{\cal O}(Z-\a X,-X)~.
\end{equation}
By scaling arguments, it is straightforward to see that ${\cal E}(\vec{n})$ transform as a 4D primary operator located at spatial infinity $i^0$, the `corner.'  While the ANEC operator is non-local from the point of view of the bulk (as evident by the $u$-integral in its definition~\eqref{equ:ANE-def-x}), what's important for us is that it is local with respect to the celestial sphere.

Amusingly, while it was already observed  in~\cite{Hofman:2008ar} that the energy detector has boost weight three, the notion of treating these objects correlators in 2D CFT was dismissed in the same breath. Our aim here is to remedy this, by connecting their construction to that of CCFT. In this section, we are interested specifically in the generalizations of the stress tensor light-ray operators that appear in the conformal collider literature, their symmetry algebras, and demonstrating that they are celestial primaries.

\subsection{Light-Ray Operators and Their Algebras}\label{sec:detector}

We will start by setting up our conventions for a variety of light-ray operators that appear in the conformal collider literature.  First, it is useful to be able to change between coordinate systems that map operators at null infinity to a hyperplane through the origin, following~\cite{Hofman:2008ar}. Starting from flat-Bondi coordinates $\{u,r,z,\bz\}$ \cite{Dumitrescu:2015fej}
\be
x^\mu=\frac{1}{2}\Big(u+r(1+z\bz), r(z+\bz), ir(\bz-z), -u+r(1-z\bz)\Big)~~,
\label{equ:Xmu-Bondi}
\ee
the Minkowski metric takes the form
\be
ds^2~=~-\,dudr+r^2\, dzd\bz~~.
\label{equ:Minkmetric}
\ee
It is useful to write~\eqref{equ:Xmu-Bondi} in terms of the lightcone components
\be
x^+~=~r~~,~~x^-~=~u+rz\bz~~,~~x^1+ix^2~=~r\,z~~,
\label{equ:xpm}
\ee
where $x^{\pm}=x^0\pm x^3$. Following~\cite{Hofman:2008ar} let us introduce the following `inverse' light-cone coordinates $\{y^+,y^-,y^1,y^2\}$ such that
\begin{equation}
    y^+ ~=~ -\frac{1}{x^+} ~~,~~
    y^- ~=~ x^- - \frac{x_1^2+x_2^2}{x^+} ~~,~~
    y^1 ~=~ \frac{x^1}{x^+} ~~,~~
    y^2 ~=~ \frac{x^2}{x^+}~~.
    \label{equ:ymap}
\end{equation}
Plugging in~\eqref{equ:xpm}, we see that
the limit $r\rightarrow \infty$ corresponding to null infinity is mapped to the hyperplane $y^+=0$ under the conformal transformation \eqref{equ:ymap}. Explicitly
\be
y^+~=~-\frac{1}{r}~~,~~y^-~=~u~~,~~ y^1+iy^2~=~z~.
\ee 
Since this coordinate transformation is a 4D conformal transformation, we can readily go back and forth between expressions for our operators in either form. The $y$ expressions have the advantage that they avoid an explicit $r\rightarrow\infty$ limit, and often appear in the conformal collider literature.

Let us now highlight some of the light-ray-supported operators that have been featured in the conformal collider literature,  before showing how these can be recast as celestial operators in the following subsection.

\paragraph{BMS Generators}

Let's start with the example most closely linked to the asymptotic symmetry analysis so far: the BMS algebra identified by Cordova and Shao~\cite{Cordova:2018ygx}. Given the assumptions of micro-causality, unitarity, Poincar\'e invariance, and closure of the commutators of light-ray operators constructed from local conserved currents, they were able to show that, for any such CFT, one can realize the BMS algebra in terms of light-ray-supported operators constructed from the stress tensor. A similar analog for the `large' U(1) gauge charges was also identified, which corresponds to the matter modes sourcing electromagnetic memory~\cite{Susskind:2015hpa,Pasterski:2015zua}.  Explicitly, the supertranslations are generated by an angular smearing of the ANEC operator, which in the $y$ coordinates (and for generic bulk dimension $d$) takes the form
\begin{equation}
        {\cal T}(f) ~\equiv~ \int\,d^{d-2}y^{\bot}\,f(y^{\bot})\,{\cal E}(y^{\bot})~~,~~   {\cal E}(y^\perp) ~=~ 2\,\int_{-\infty}^{+\infty}\,dy^-\,T_{--}(y^-,y^+=0,y^\perp)~~,
        \label{equ:Tf}
    \end{equation}
    where $f(y^{\bot})$ is an arbitrary function. Meanwhile, the superrotation generator is defined as 
    \begin{equation}
        {\cal R}(Y^A) ~\equiv~ \int\,d^{d-2}y^{\bot}\,Y^A(y^{\bot})\,{\cal N}_A(y^{\bot}) + \frac{1}{d-2}\,\int\,d^{d-2}y^{\bot}\,\pa_AY^A(y^{\bot})\,{\cal K}(y^{\bot})~~,
    \end{equation}
where
\begin{equation}
    {\cal N}_A(y^{\bot}) = \int_{-\infty}^{+\infty} dy^-\,T_{-A}(y^-,y^+=0,y^{\bot})~~,~~
    {\cal K}(y^{\bot}) = \int_{-\infty}^{+\infty}dy^-\,y^-\,T_{--}(y^-,y^+=0,y^{\bot})~~,
\end{equation}
and, again, all of these operators are parallel and live on the same light sheet.  While the main power comes from the bottom-up arguments and minimal assumptions the authors  of~\cite{Cordova:2018ygx} are able to start from, they also explicitly demonstrate this algebra holds for correlators in a free scalar theory. Since then, the Cordova-Shao light-ray algebra has been explicitly checked perturbatively in generic 4D QFTs of massless particles with integer spin in \cite{Gonzo:2020xza}.  This fits nicely with the picture that the extrapolate dictionary in celestial CFT treats all massless $\cal S$-matrix elements as correlators of operators on past and future null infinity, and that we are interested in perturbation theory around the free CFT.

Rewriting these generators in terms of $\{u,r,z, \bz\}$ coordinates for $d=4$, it is straightforward to see that these operators are none other than the (outgoing) hard charges for the corresponding asymptotic symmetries $Q_H^{{\cal I}^+}(f,Y)$~\cite{He:2014laa,Lysov:2014csa}.  Similar expressions can be constructed from their $\mathcal{I}^-$ counterparts, which give the remaining part of the hard charge for a purely massless theory (i.e. $Q_H^{\pm}=Q_H^{{\cal I}^\pm}$). Once we couple to gravity, the Ward identity
\be
\langle out |Q^+(f,Y)\mathcal{S}-\mathcal{S} Q^-(f,Y)|in\rangle=0~~,~~Q^\pm(f,Y)=Q_S^\pm(f,Y)+Q_H^\pm(f,Y)~~,
\label{eq:ward}
\ee
where $Q_S$ is a soft charge that inhomogeneously shifts the vacuum state, follows in principle from an antipodal matching of the charges across spatial infinity, and in practice from the leading and subleading soft graviton theorems. For the supertranslations and superrotations, we can split the soft and hard parts so that the soft, hard graviton, and hard matter contributions $\{ Q_S, Q_H^{grav},Q_H^{M}\}$ each form a representation of the corresponding asymptotic symmetry algebra and are mutually commuting.\footnote{For details for how to arrange such a hard and soft splitting in pure gravity see~\cite{Donnay:2021wrk}. For how this helps us understand loop-level corrections see~\cite{Pasterski:2022djr,Donnay:2022hkf}. } Given this structure one can expect that if we `decouple' gravity we are left with the matter sector obeying the same symmetries. However, it is perhaps still surprising that the minimal assumptions that went into the~\cite{Cordova:2018ygx} derivation are enough for the CFT to be `ready' to couple to gravity.

\paragraph{Virasoro Extension}
While the Cordova and Shao result makes more direct contact with the `soft theorem = Ward Identity' results of Strominger et al., due to their shared interest in examining local enhancements of Poincar\'e symmetry, \cite{Cordova:2018ygx} itself is actually an extension of results by Casini et al.~\cite{Casini:2017roe}. The analysis there was in an information-theoretic context, and the authors were studying modular Hamiltonians for regions with null horizons~\cite{Casini:2017roe}.  In practice, they were interested in the algebra of parallel ${\cal E}(y^{\bot})$ and ${\cal K}(y^{\bot})$ inserted on the same light sheet, but they also identified an extension.

Namely, the argument that these commutators involve a line integral of $T_{--}$ also applies to more general light-ray-supported operators of the form\footnote{Note that this operator is ill-defined for sufficiently high $n$. Meanwhile, for $n<-2$ there is a pole at $y^-=0$. As discussed in \cite{Belin:2020lsr}, one can define \begin{equation}
    {\cal L}_{n}(y^{\bot}) ~=~  \lim_{\delta\to 0^+}\int_{-\infty}^{+\infty}\,dy^-\,e^{i\delta y^-}\,(y^-)^{n+2} \,T_{--}(y^-,y^+=0,y^{\bot})
    \label{equ:Ln-high-n}
\end{equation}
and for $n<-2$, one must choose a $i\epsilon$-prescription for the pole at $y^-=0$. 
}
\begin{equation}
    {\cal L}_{n}(y^{\bot}) ~=~  \int_{-\infty}^{+\infty}\,dy^-\,(y^-)^{n+2} \,T_{--}(y^-,y^+=0,y^{\bot})~~.
    \label{equ:Ln}
\end{equation}
By matching dimensions and twists, assuming the commutator has to be proportional to a delta function in the transverse directions, and applying Jacobi identities, they argued these light-ray operators obeyed a Virasoro algebra 
\begin{equation}
    [{\cal L}_n(y^{\bot}),{\cal L}_m(y'^{\bot})] ~=~ -i\,(m-n)\,\delta^{(d-2)}(y^{\bot}-y'^{\bot})\,{\cal L}_{m+n+1}(y'^{\bot}) ~~,
    \label{eq:Virasoso}
\end{equation}
where any central charge would be UV divergent on dimensional grounds \cite{Wall:2011hj}.  More recently, Belin et al.~\cite{Belin:2020lsr} revisited this algebra in explicit correlation functions, focusing on $d=4$. They showed that it breaks down due to additional terms with support at finite spacelike separation, which they attribute to the light-ray integrals of the operators being ill-defined. While one can check that we can realize this algebra for the free boson at the level of the creation and annihilation operator algebra, understanding the convergence of various $u$-moments of our fields is important for determining how to handle conformally soft modes in CCFT and, more generally, early and late-$u$ boundary conditions we can impose on our phase space that will still be consistent with the unitarity of the $\cal S$-matrix.

\paragraph{Detector Operators}

Finally, there is another generalization of the ANEC operator that is natural from the point of view of the bulk CFT: the so-called `detector operators' studied, for example, in the context of weakly coupled field theories in~\cite{Caron-Huot:2022eqs}. While, as we discussed above, the algebra of ANEC operators with their superrotation counterpart closes, the 
OPE of two ANEC operators includes more general non-local operators built from products of operators separately smeared along generators of null infinity~\cite{Hofman:2008ar,Kologlu:2019mfz,Chang:2020qpj}.  There exists a special class of light-ray operators that are a 
 continuous spin generalization of the ANEC operator~\cite{Kravchuk:2018htv}
\begin{equation}
    {\cal E}_J ~=~ 2\,{\bf L}[{\cal O}_J(X_{i^0},Z_{\infty})] ~~,
\end{equation}
where for a free scalar, ${\cal O}_J$ is an analytic continuation in $J$ of the double trace operator
\begin{equation}
    {\cal O}_J ~=~ N_J\,:\phi(x)(z\cdot\pa)^J\phi(x): ~+~ (z\cdot\pa)(\cdots) ~.
\end{equation}
Here the null vector $z$ picks out the $\p_u$ derivative along the corresponding generator of null infinity. After some integration by parts, we see that in Bondi coordinates we can write 
\begin{equation}
     {\cal E}_J(z,\bz) ~=~ \lim_{r\to\infty}\,r^2\,\int\,du\,:\pa_u\phi\,\pa_u^{J-1}\phi:(u,r,z,\bz)~~,
     \label{eq:calE}
\end{equation}
where we emphasize that $J$ need not be a positive integer. As we will see explicitly in section~\ref{sec:LRasCelestialPrimary}, ${\cal E}_J(z,\bz)$ is a celestial primary with $\Delta_{2D}=J+1$ and $J_{2D}=0$.

This takes a somewhat cleaner form in momentum space, as we will now outline. The radiative data for a scalar field is in the $\mathcal{O}(r^{-1})$ mode 
\begin{equation}
    \phi(u,r,z,\bz) ~=~ \frac{\phi^{(0)}(u,z,\bz)}{r} ~+~ \cdots~~.
\end{equation}
Note that in the definition of $ {\cal E}_J(z,\bz)$, due to the large-$r$ limit, we only keep the leading term involving $\phi^{(0)}$. For the free theory we can use the mode expansion 
\begin{equation}
    \phi^{(0)}(u,z,\bz) ~=~ \frac{i}{(2\pi)^2}\, \int_0^{\infty}\,d\omega\,\Big[ a^{\dagger}(\omega,z,\bz)\,e^{i\omega u} - a(\omega,z,\bz)\,e^{-i\omega u}\Big]~~,
    \label{eq:phi0}
\end{equation}
and we see that
\begin{equation}
    \pa^{J-1}_u\,\phi^{(0)}(u,r,z,\bz) ~=~ \frac{i}{(2\pi)^2}\, \int_0^{\infty}\,d\omega\,(i\omega)^{J-1}\,\Big[ a^{\dagger}(\omega,z,\bz)\,e^{i\omega u} + (-1)^{J} a(\omega,z,\bz)\,e^{-i\omega u}\Big]~~.
\end{equation}
 Plugging these into~\eqref{eq:calE} we thus have 
\begin{equation}
    \begin{split}
        {\cal E}_J(z,\bz) ~=&~ \lim_{r\to\infty}\,r^2\,\int\,du\,\frac{1}{r^2}\,:\pa_u\phi^{(0)}(u,r,z,\bz)\,\pa_u^{J-1}\phi^{(0)}(u,r,z,\bz):\\
        ~\propto &~ \int_0^{\infty}\,d\omega\,\omega^{J}\,:a^{\dagger}(\omega,z,\bz)a(\omega,z,\bz):~,
    \end{split}
\end{equation}
where we have used the identity \eqref{equ:exp-Fourier}. As an aside, we can also see from this expression that this should have positive expectation values for real $J$ (the authors of~\cite{Kravchuk:2018htv} present a proof that this holds beyond the free limit). These modes measure a power of the energy deposited in each direction, while they themselves add zero energy to the out state, and are thus part of the `soft' matter sector.

It is straightforward to evaluate the commutation relations for these modes of the free scalar. Given the canonical commutation relation
\begin{equation}\label{eq:adaggera}
    [a(\omega,z,\bz), a^{\dagger}(\omega',w,\bw)] ~=~ (2\pi)^3\,\frac{2}{\omega}\,\delta(\omega-\omega')\,\delta^{(2)}(z-w)~~,
\end{equation}
we can calculate the commutators of 
${\cal E}_J(z,\bz)$'s to be
\begin{equation}\scalemath{.92}{
    \begin{aligned}
        \Big[{\cal E}_{J_1}(z,\bz) , {\cal E}_{J_2}(w,\bw)\Big] =&  \int_0^{\infty}\,d\omega_1\,\omega_1^{J_1}\,\int_0^{\infty}\,d\omega_2\,\omega_2^{J_2}\,\Bigg\{ a^{\dagger}(\omega_2,w,\bw)\Big[a^{\dagger}(\omega_1,z,\bz),a(\omega_2,w,\bw)\Big]a(\omega_1,z,\bz)  \\
        &\qquad~+~ a^{\dagger}(\omega_1,z,\bz)\Big[a(\omega_1,z,\bz),a^{\dagger}(\omega_2,w,\bw)\Big]a(\omega_2,w,\bw) \Bigg\}\\
        ~=&~ 0~~.
    \end{aligned}}
\end{equation}
We see that, unlike the ${\cal L}_n$ modes, these operators commute.

\vspace{1em}

It is natural to ask about an asymptotic symmetry interpretation of these generalizations, and we will find in section~\ref{sec:tower} that we need a combination of operators of the ${\cal L}_n$-type (higher $u$-moments) and ${\cal E}_J$-type (analytic continuations of $\p_u^J$) to reproduce the celestial $w_{1+\infty}$ including matter. Moreover, we will see that the notion of the matter sector decoupling to a representation of the same algebra is not something we should take for granted.  But first, let us now show that these are all indeed celestial primaries!

\subsection{
Light-Ray Operators as Celestial Primaries}\label{sec:LRasCelestialPrimary}

Now that we have provided an overview of the light-ray-supported operators that often appear in the conformal collider literature, the goal of this section is to rephrase these operators in a manifestly celestial language. While this won't be the first place mentioning the boost weights of these operators (see e.g.~\cite{Hofman:2008ar,Kologlu:2019mfz}), the main point here is to emphasize that organizing operators in terms of `boost selection rules' a la~\cite{Kologlu:2019mfz} is indeed what underpins the celestial framework. We'll start with the power counting presentation that might be more familiar from the conformal collider literature, before presenting things in the CCFT language.

As presented in more detail in Appendix~\ref{app:4Dcoords}, the manner in which the CFT$_4$ generators get conjugated by the coordinate map~\eqref{equ:ymap} is such that the boost in $x^+$, $x^-$ plane becomes the dilatation in $y$ and vice versa
\be
\Delta_{2D}:=M^{03}_x = D_y~~,~~\Delta_{4D}:= D_x=M^{03}_y ~~.
\ee
We can keep track of this via the power counting
\be
x^\mu\mapsto \lambda x^\mu ~~\Leftrightarrow~~ (y^+,y^-,y^\perp)\mapsto (\lambda^{-1}y^+,\lambda y^-, y^\perp)~~,
\ee
and similarly with $x$ and $y$ exchanged. The scaling dimensions of the Bondi coordinates are listed in Table~\ref{tab:list-coord-4D2D} for reference. In particular, if we started with an operator with definite dilatation weight at the origin, and instead insert it at the $u=0$ cut of $\mathcal{I}^+$, that state will now have definite collinear boost weight. 

\begin{table}[htp!]
    \centering
    \begin{tabular}{c|c|c|c}
       Coordinate  &  $\Delta_{\rm 4D}$ & $\Delta_{\rm 2D}=J^{03}$ & $J_{\rm 2D}=J^{12}$  \\\hline
       $u$ & $-1$ & $-1$ & 0 \\
       $r$ & $-1$ & 1 & 0 \\
       $z$ & 0 & $-1$ & $-1$ \\
       $\bz$ & 0 & $-1$ & 1  \\
    \end{tabular}
 \caption{Various scaling dimensions for our Bondi coordinates.}
    \label{tab:list-coord-4D2D}
\end{table}

Let's now translate things into a manifestly celestial language. The operators we are interested in are designed to be conformal (quasi)-primaries living on the celestial sphere
\be
{\cal O}^\pm_{\Delta,J}\left(\frac{a w+b}{cw+d},\frac{{\bar a} \bw+{\bar b}}{{\bar c}\bw+{\bar d}}\right)=(cw+d)^{\Delta+J}({\bar c}\bw+{\bar d})^{\Delta-J}{\cal O}^\pm_{\Delta,J}(w,\bw)~~.
\label{eq:Omobius}
\ee
For the single-particle states, we can construct the corresponding celestial operators using the natural inner product on solutions of the free wave equation
 \be
 {\cal O}^\pm_{\Delta}=i(\hat{O}(x),\Phi_{\Delta^*,-J}(x_\mp;w,\bw))_\Sigma ~~,
 \ee
 where $x_\pm^0=x^0\mp i\epsilon$ picks out the appropriate outgoing/incoming modes.  The transformation property~\eqref{eq:Omobius} then is guaranteed by the covariance of the conformal primary wavefunctions~\cite{Pasterski:2017kqt}
\begin{equation}\label{Defgenprim}
    \badat{2}
\Phi^{s}_{\Delta,J}\Big(\Lambda^{\mu}_{~\nu} x^\nu;\frac{a w+b}{cw+d},\frac{{\bar a} \bw+{\bar b}}{{\bar c}\bw+{\bar d}}\Big)=(cw+d)^{\Delta+J}({\bar c}\bw+{\bar d})^{\Delta-J}D_s(\Lambda)\Phi^{s}_{\Delta,J}(x^\mu;w,\bw)\,
\eadat
\end{equation}
under Lorentz transformations. Acting on the vacuum with one of these operators creates a state with definite $\mathrm{SL}(2,\mathbb{C})$ weights
 \be |\Delta,J\rangle\equiv \O_{\Delta,J}(0) |0\rangle ~~.
\ee
For massless scattering, this state can be reached by starting from the momentum eigenstates with momenta aligned in the same direction as the reference point ($w=z$) and performing a Mellin transform in the frequency 
\be
\langle\Delta,J| \propto \int_0^\infty d\omega \omega^{\Delta-1}\langle\omega,z=0|\propto \Gamma(\Delta)\lim\limits_{r\rightarrow\infty}\int_{-\infty}^{\infty} du (u-i\epsilon)^{-\Delta}\langle 0|\phi(u,r,z,\bz)~~.
\ee
Meanwhile, from the point of view of the celestial extrapolate dictionary, these states are prepared by taking $u$-moments of operators smeared along generators of null infinity.

The general multi-particle case is treated systematically in the companion paper~\cite{kp}. For the particular conformally soft modes relevant here, it is straightforward to check that they are celestial primaries given the Lorentz transformation properties of the stress tensor and scalar field. First we note that the Lorentz transformation acting on $x^{\mu}\mapsto \Lambda^{\mu}{}_{\nu}x^{\nu}$ in~\eqref{Defgenprim} takes the explicit form
\begin{equation}\resizebox{0.9\textwidth}{!}{$%
    \Lambda^{\mu}{}_{\nu} ~=~ \frac{1}{2}\begin{pmatrix}
    a\ba+b\bb+c\bc+d\bd & a\bb+\ba b+\bc d+ c\bd & i(a\bb-\ba b+c\bd - \bc d) & b\bb-a\ba-c\bc +d\bd\\
    a\bc+\ba c+b\bd + \bb d & a\bd+\ba d+b\bc+\bb c & i(a\bd-\ba d - b\bc+\bb c) & b\bd+\bb d-a\bc-\ba c \\
    i(\ba c-a\bc-b\bd+d\bb) & i(\ba d-a\bd - b\bc + \bb c) & a\bd+\ba d-b\bc-\bb c & i(a\bc-\ba c-b\bd+\bb d)\\
    c\bc+d\bd-a\ba-b\bb & c\bd+\bc d- a\bb - \ba d & i(\ba b- a\bb+c\bd-\bc d) & a\ba-b\bb-c\bc +d\bd
    \end{pmatrix}$}.
\end{equation}
In particular, this implies that our Bondi coordinates transform as follows\footnote{A detailed exposition of this calculation can be found in \cite{Oblak:2015qia}. Note that we choose a different parametrization in~\eqref{equ:Xmu-Bondi} as compared with \cite{Oblak:2015qia}. In our case, we have
\begin{equation}
    r= x^0+x^3~~,~~ u=x^0-x^3-\frac{(x^1)^2+(x^2)^2}{x^0+x^3} ~~,~~ z=\frac{x^1+ix^2}{x^0+x^3} ~~,~~ \bz=\frac{x^1-ix^2}{x^0+x^3}
\end{equation}
and the transformation laws of the Bondi coordinates in (\ref{equ:Lorentz-Bondi}) are much simpler than the ones in \cite{Oblak:2015qia}.}
\begin{equation}\scalemath{.95}{
    \begin{aligned}
        r' ~=&~ r\,|cz+d|^2+|c|^2u ~=~ r\,|cz+d|^2 ~+~ {\cal O}(1) ~~,\\
        u' ~=&~ \frac{r\,u}{r|cz+d|^2+|c|^2\,u} ~=~\frac{u}{|cz+d|^2} ~+~ {\cal O}(1/r) ~~,\\
        z' ~=&~ \frac{r(az+b)(\bc\bz+\bd)+a\bc u}{r(cz+d)(\bc\bz+\bd)+|c|^2u} ~=~\frac{az+b}{cz+d} ~+~ {\cal O}(1/r) ~~,\\
        \bz' ~=&~ \frac{r(cz+d)(\ba\bz+\bb)+\ba c u}{r(cz+d)(\bc\bz+\bd)+|c|^2u} ~=~\frac{\ba\bz+\bb}{\bc\bz+\bd} ~+~ {\cal O}(1/r) ~~.
    \end{aligned}}
    \label{equ:Lorentz-Bondi}
\end{equation}
Writing all the light-ray operators defined in the previous section in Bondi coordinates gives us
\be\scalemath{.97}{\badat{3}
    {\cal E}(z,\bz) ~=&~ \lim_{r\to\infty}\,r^2\,\int\,du\,T_{uu}(u,r,z,\bz)~, \\
    {\cal K}(z,\bz) ~=&~ \lim_{r\to\infty}\,r^2\,\int\,du\,u\,T_{uu}(u,r,z,\bz)~, \\
    {\cal N}_A(z,\bz) ~=&~ \lim_{r\to\infty}\,r^2\,\int\,du\,T_{uA}(u,r,z,\bz)~, \\
    {\cal J}^a(z,\bz) ~=&~ \lim_{r\to\infty}\,r^2\,\int\,du\,j_u^a(u,r,z,\bz)~, \\
    {\cal L}_n(z,\bz) ~=&~ \lim_{r\to\infty}\,r^2\,\int\,du\,u^{n+2}\,T_{uu}(u,r,z,\bz)~, \\
    {\cal E}_J(z,\bz) ~=&~ \lim_{r\to\infty}\,r^2\,\int\,du\,:\pa_u\phi\,\pa_u^{J-1}\phi:(u,r,z,\bz)~,
\eadat}\ee
where, in the fourth line, $a$ is an adjoint index for a non-Abelian large gauge symmetry.
Taking the ANEC as a concrete example, we see that the 4D Lorentz transformation laws imply\footnote{Note that we only care about the leading term in the radial expansion in the $r\to\infty$ limit.}
\begin{equation}
    \begin{split}
    {\cal E}&\left(z'=\frac{az+b}{cz+d},\bz'=\frac{\bar{a}\bz+\bar{b}}{\bar{c}\bz+\bar{d}}\right) ~=~ \lim_{r'\to\infty}\,r'^2\,\int\,du'\,T_{u'u'}(u',r',z',\bz') \\
    &~=~ |cz+d|^{4}\,\lim_{r\to\infty}\,r^2\,\int\,\frac{du}{|cz+d|^2} \,\left(\frac{\pa u}{\pa u'}\right)^2\,T_{uu} ~=~ (cz+d)^{3}\,(\bar{c}\bz+\bar{d})^3\,{\cal E}(z,\bz) ~,
    \end{split}
\end{equation}
which means ${\cal E}(z,\bz) $ is a $\Delta=3$, $J=0$ primary on the celestial sphere (where for convenience we've suppressed the 2D subscript on the weights and spins here). Similarly, we have
\be\scalemath{0.95}{\begin{array}{rclclclc lcl}
    {\cal K}(z',\bz') &=&~ (cz+d)^{2}\,(\bar{c}\bz+\bar{d})^2\, {\cal K}(z,\bz) ~~&\leftrightarrow&~~ \Delta_{{\cal K}}~&=& 2, ~~&~~ J_{\cal K}~&=&0~, \\
    {\cal N}_z(z',\bz') &=&~  (cz+d)^{4}\,(\bar{c}\bz+\bar{d})^2\,{\cal N}_z(z,\bz) ~~&\leftrightarrow&~~ \Delta_{{\cal N}_z}~&=& 3, ~~&~~ J_{{\cal N}_z}~&=& 1~,\\ 
    {\cal N}_{\bz}(z',\bz') &=&~  (cz+d)^{2}\,(\bar{c}\bz+\bar{d})^4\,{\cal N}_{\bz}(z,\bz) ~~&\leftrightarrow&~~ \Delta_{{\cal N}_{\bz}}&=& 3, ~~&~~ J_{{\cal N}_{\bz}}&=& -1~,\\ 
    {\cal J}^a(z',\bz') &=&~ (cz+d)^{2}\,(\bar{c}\bz+\bar{d})^2\, {\cal J}^a(z,\bz) ~~&\leftrightarrow&~~ \Delta_{ {\cal J}^a}&=& 2, ~~&~~ J_{ {\cal J}^a}&=& 0~,\\ 
    {\cal L}_n(z',\bz') &=&~ (cz+d)^{1-n}\,(\bar{c}\bz+\bar{d})^{1-n}\, {\cal L}_n(z,\bz) ~~&\leftrightarrow&~~ \Delta_{{\cal L}_n}&=& 1-n, ~~&~~ J_{{\cal L}_n }&=& 0~,\\
     {\cal E}_J(z',\bz') &=&~ (cz+d)^{J+1}\,(\bc\bz+\bd)^{J+1}\,{\cal E}_J(z,\bz) ~~&\leftrightarrow&~~ \Delta_{{\cal E}_J}&=& 1+J, ~~&~~ J_{{\cal E}_J }&=& 0~.
\end{array}}\ee
Thus, we see that all the light-ray operators we have discussed so far are indeed celestial primaries and obtained their conformal weights, which are summarized in Table \ref{tab:list-lightrayop-4D2D}. Note that although ${\cal E}_J(z,\bz)$ and ${\cal L}_{-n}$ have the same conformal weights, they are not the same primaries, as one can see clearly from their mode expansions in the free limit. While most of the literature on celestial amplitudes focuses on the gauge particles coupling to these operators, we note that the matter modes of weight $\Delta=s+1$ here have appeared as the hard charges in celestial Ward identities. The goal of the next section is to show how we can use this perspective to extend these constructions!

\begin{table}[htp!]
    \centering
    \scalemath{.92}{
    \begin{tabular}{l|c|c|c}
     ~~~~~~~~~~~~~~~~~Light-ray  Operators  & $\Delta_{\rm 2D}=J^{03}(x)= \Delta_{\rm 4D}(y)$ & $J_{2D}=J^{12}$ & $J^{03}(y)=\Delta_{4D}(x)$  \\\hline
       ${\cal E}(y^{\bot}) \equiv \int_{-\infty}^{+\infty}dy^-T_{--}(y^-,y^+=0,y^{\bot})$  &  3 &  0 & 1 \\
       ${\cal K}(y^{\bot}) \equiv \int_{-\infty}^{+\infty}dy^-y^-\,T_{--}(y^-,y^+=0,y^{\bot}) $  &  2 &  0 & 0 \\
       ${\cal N}_A(y^{\bot}) \equiv \int_{-\infty}^{+\infty}dy^-T_{-A}(y^-,y^+=0,y^{\bot}) $  &  3 &  $\pm 1$ & 1 \\
       $ {\cal J}^a(y^{\bot}) \equiv \int_{-\infty}^{+\infty}dy^-j^a_{-}(y^-,y^+=0,y^{\bot}) $  &  2 &  0 & 0 \\
       ${\cal L}_{n}(y^{\bot}) \equiv \int_{-\infty}^{+\infty}dy^-(y^-)^{n+2} T_{--}(y^-,y^+=0,y^{\bot})$  &  $1-n$ &  0 & $-1-n$ \\
    \end{tabular}}
    \caption{Summary of the light-ray operators we've encountered, as well as their scaling dimensions for the various symmetry transformations we encounter in the $x$ and $y$ coordinate systems and how these translate to their celestial spectrum.
    }
    \label{tab:list-lightrayop-4D2D}
\end{table}

\section{A Tower of Symmetries}\label{sec:tower}

While the light-ray operators studied in~\cite{Cordova:2018ygx} are known to be connected to BMS and large gauge asymptotic symmetries, the symmetry structure in CCFT doesn't stop there. Indeed, over the past couple of years there has been much excitement in the celestial literature~\cite{Guevara:2021abz,Himwich:2021dau,Adamo:2021lrv,Freidel:2021ytz,Donnay:2022sdg} (see also~\cite{Hamada:2018vrw,Compere:2019odm,Mao:2020vgh}) surrounding an infinite tower of conformally soft theorems~\cite{Guevara:2019ypd} which, for gravity, generate a $w_{1+\infty}$ symmetry~\cite{Strominger:2021mtt}.

We now turn to the interesting question of how these recent developments in celestial holography point to a larger class of detector-like light-ray operators. In section \ref{sec:winftyalg}, we will briefly review how the $w_{1+\infty}$ symmetry arises from two different perspectives: starting from the celestial OPE, following~\cite{Guevara:2021abz} and via the gravitational phase space, following~\cite{Freidel:2021ytz}. Our main results are presented in section \ref{sec:algresult}, where we show how to realize $w_{1+\infty}$ in a 4D complex scalar theory coupled to gravity. This realization involves conformally soft `multi-particle' operators (i.e. higher order in the fundamental fields at null infinity) which are non-local on null infinity and naturally extend the library of light-ray operators that may be of interest to the conformal collider literature.  In the process, we will see how having a massless spin-2 mode affects the symmetry algebras we expect, and how the decoupling limit for our $w_{1+\infty}$ generators differs from that of the BMS subalgebra.

\subsection{Identifying Symmetries from the OPE vs from Phase Space}\label{sec:winftyalg}

Let us now review two complementary constructions of the $w_{1+\infty}$ symmetry of gravity.  The first~\cite{Guevara:2021abz,Strominger:2021mtt} starts from collinear limits of positive helicity gravitons amplitudes and is how this algebra was first identified within CCFT. The second approach lifts the self-dual restriction~\cite{Freidel:2021ytz}, by constructing the algebra in terms of multi-graviton operators.

\paragraph{Symmetries from the Celestial OPE} 

The starting point of celestial CFT is that the Lorentz group of the bulk acts as conformal isometries of the celestial sphere. As such, $\cal S$-matrix elements written in a basis of boost eigenstates transform as correlation functions of (quasi)-primaries in a 2D CFT. If we want to study this 2D CFT, it is natural to start with its OPE. This can be extracted from collinear limits of scattering~\cite{Fan:2019emx,Pate:2019lpp}. Upon applying a Mellin transform to the collinear splitting function in Einstein gravity, we get the graviton OPE
 \be\badat{3}\label{ggope}
\mathcal{O}_{\Delta_1,+2}(z_1,\bz_1)\mathcal{O}_{\Delta_2,+2}(z_2,\bz_2)&\sim-\frac{\kappa}{2}\frac{\bz_{12}}{z_{12}}B(\Delta_1-1,\Delta_2-1)\mathcal{O}_{\Delta_1+\Delta_2,+2}(z_2,\bz_2)+...~,\\
\mathcal{O}_{\Delta_1,+2}(z_1,\bz_1)\mathcal{O}_{\Delta_2,-2}(z_2,\bz_2)&\sim-\frac{\kappa}{2}\frac{\bz_{12}}{z_{12}}B(\Delta_1-1,\Delta_2+3)\mathcal{O}_{\Delta_1+\Delta_2,-2}(z_2,\bz_2) \\
&~~~-~ \frac{\kappa}{2}\frac{z_{12}}{\bz_{12}}B(\Delta_1+3,\Delta_2-1)\mathcal{O}_{\Delta_1+\Delta_2,+2}(z_2,\bz_2)+...~,
\eadat\ee
where the omitted terms are subleading in the (complexified) collinear limit. In particular, we see that the OPE coefficients are proportional to beta functions which have poles for a semi-infinite tower of weights. At the leading and subleading orders, these residues are precisely the conformally soft theorems~\cite{Cheung:2016iub,Pate:2019mfs,Puhm:2019zbl,Adamo:2019ipt} corresponding to supertranslation and superrotation symmetries~\cite{Donnay:2018neh,Donnay:2020guq}. More generally we can consider the residues
\begin{equation}
    H^k(z,\bz) ~:=~ \lim_{\epsilon\to 0}\,\epsilon\,{\cal O}_{k+\epsilon,2}(z,\bz),~~~\Delta=k ~=~2, 1,0,-1,\dots~
\end{equation}
for positive helicity gravitons.
The authors of~\cite{Guevara:2021abz} noticed that the OPE closes on these modes. These conformal dimensions correspond to $SL(2,\mathbb{R})_R$ weights
\be
\bar h=\frac{1}{2}(\Delta-J)=0,-1,-2, \ldots
\ee
such that these states have primary descendants~\cite{Pasterski:2021fjn,Pasterski:2021dqe}. In a radially quantized 2D CFT these primary descendants would be null states we would expect to decouple. Examining the leading term in an expansion of the wavefunctions as $\Delta\rightarrow k$ one indeed sees that these states would be in shortened multiplets for $k\le-2$. For such shortened multiplets we can write the mode expansion
\begin{equation}
    H^k(z,\bz) ~=~  \sum_{m=\frac{k-2}{2}}^{\frac{2-k}{2}}\,\bz^{-\frac{k-2}{2}-m}\,H^k_m(z) ~,
    \label{eq:hkmode}
\end{equation} 
which \cite{Guevara:2021abz} observed is consistent with the leading OPE~\eqref{ggope}. Attempting to treat this system like a 2D CFT,~\cite{Guevara:2021abz} then evaluated the (complexified) radial quantization commutator
\begin{equation}
    [A,B](z)~=~\frac{1}{2\pi i}\,\oint_z\, dw\, A(w)B(z)
\end{equation}
to obtain an algebra for these modes
\begin{equation}\scalemath{.95}{
    \Big[ H^k_m, H^l_n \Big] = -\frac{\kappa}{2}\Big[ n(2-k)-m(2-l)\Big]\frac{(\frac{2-k}{2}-m+\frac{2-l}{2}-n-1)!(\frac{2-k}{2}+m+\frac{2-l}{2}+n-1)!}{(\frac{2-k}{2}-m)!(\frac{2-l}{2}-n)!(\frac{2-k}{2}+m)!(\frac{2-l}{2}+n)!}\,H_{m+n}^{k+l}\,}~.
    \label{equ:H-commutation}
\end{equation}
This expression simplifies drastically with a simple redefinition 
\begin{equation}
    w^p_n ~=~ \frac{1}{\kappa}\,(p-n-1)!(p+n-1)!\,H_n^{-2p+4}~,
\end{equation}
after which \eqref{equ:H-commutation} becomes
\begin{equation}
    \Big[w^p_n, w^q_m\Big](z) ~=~ \Big[ n(q-1) - m(p-1)\Big]\,w^{p+q-2}_{m+n}(z)~,
    \label{eq:wloop}
\end{equation}
where $p=1,\frac{3}{2},2,\frac{5}{2}...$ and $1-p\le m\le p-1$
and we see that the (complexified) radial quantization brackets of the positive helicity gravitons generate the wedge subalgebra of the loop algebra of $w_{1+\infty}$. This symmetry can be traced back to the $w_{1+\infty}$ symmetry of self-dual gravity~\cite{Adamo:2021lrv}, and these explorations have spurred a fascinating merger with results from both twistor theory and twisted holography~\cite{Costello:2022wso,Costello:2022upu,Costello:2022jpg,Bu:2022iak}. One downside of the original derivation is that it started from tree-level amplitudes, so it was not immediately obvious how these results generalized to loop level or, more importantly, beyond self-dual theories. In~\cite{Ball:2021tmb}, it was shown that this symmetry survived to all loop orders in self-dual gravity. We will now turn to a reformulation of this $w_{1+\infty}$ symmetry that doesn't require a restriction to self-dual theories.

\paragraph{Symmetries from their canonical representation on phase space}

While understanding the $w_{1+\infty}$ symmetry in terms of self-dual gravity is helpful, the derivation in terms of the OPE shown above is in Einstein gravity.  Mechanically, the choice of doing $z$ as opposed to $\bz$ contour integrals picked out the algebra of positive helicity modes. While an analogous algebra would hold for only-minus helicity operators,  it's not quite straightforward to combine the two helicities in this setup. The authors of~\cite{Freidel:2021ytz}, however, were able to reproduce the same symmetry algebra in terms of the gravitational phase space. While we can still focus on the part corresponding to the single helicity single graviton modes, the brackets are now the standard ones defined on the radiative phase space.  Their derivation further lifts the restriction to the wedge subalgebra. In keeping with their notation, we will look at what amounts to the negative helicity algebra in what follows, related by complex conjugation to the results in of~\cite{Guevara:2021abz,Strominger:2021mtt} discussed above.

Before diving into the construction of~\cite{Freidel:2021ytz}, let us briefly set up some notation.  We denote the two helicity modes of the shear tensor by ${C}(u,z,\bz)$, $\bar{C}(u,z,\bz)$ with $J=\pm 2$ respectively. These admit the following mode expansions~\cite{Strominger:2013jfa,Freidel:2021ytz}  
\begin{equation}
    \begin{split}
        C(u,z,\bz) ~=&~ \frac{i\kappa}{8\pi^2}\,\int_0^{\infty}d\omega\,\Big[ a_-^{\dagger}(\omega,z,\bz)e^{i\omega u} - a_+(\omega,z,\bz)e^{-i\omega u}\Big] ~~,\\
    \bar{C}(u,z,\bz) ~=&~ \frac{i\kappa}{8\pi^2}\,\int_0^{\infty}d\omega\,\Big[ a_+^{\dagger}(\omega,z,\bz)e^{i\omega u} - a_-(\omega,z,\bz)e^{-i\omega u}\Big] ~~,
    \end{split}
    \label{equ:mode-exp-C}
\end{equation}
in terms of the creation and annihilation operators of the graviton acting on the out state. The news tensor is related to the shear tensor by $N^{zz}=\pa_u\bar{C}$ and $N^{\bz\bz}=\pa_u C$. Their Poisson bracket
\begin{equation}
    \{ \pa_u\bar{C}(u,z,\bz),\,C(u',w,\bw)\} ~=~ \frac{\kappa^2}{2}\,\delta(u-u')\,\delta^{(2)}(z-w)~~,
\end{equation}
is equivalent to our normalization for the canonical commutation relations in~\eqref{eq:adaggera}, with an additional helicity index. In what follows we will use units where $\kappa=2$.

To compare with the celestial sphere mode expansions, we introduce the formal presentation of the Dirac delta function used in~\cite{Donnay:2021wrk}
\begin{equation}
    \delta(z-w)~=~ z^{-1}\,\sum_k\,z^k\,w^{-k}~~,~~
    \delta(\bz-\bar{w})~=~ \bz^{-1}\,\sum_k\,\bz^k\,\bar{w}^{-k}~~,~~
    \label{equ:deltafun-mode-exp}
\end{equation}
so that
\begin{equation}
    \delta^{(2)}(z-w) ~=~ \delta(z-w)\,\delta(\bz-\bar{w})~~.
\end{equation}
Rather than the truncated expansion in~\eqref{eq:hkmode}, a general conformal field of dimension $\Delta$ and spin $J$ can be written in terms of the following double mode expansion
\begin{equation}
    \phi_{\Delta,J}(z,\bz) ~=~ \sum_{k,l}\,\phi_{\Delta,J;k,l}\,z^{-\frac{\Delta+J}{2}-k}\bz^{-\frac{\Delta-J}{2}-l}~~,
    \label{equ:general-mode-exp}
\end{equation} where $k$ and $l$ run over all integers. The higher spin symmetry generators $W_s$ will have $(\Delta,J)=(3,s)$ and, as such, we will start from the general mode expansion
\begin{equation}
    W_s(z,\bz) ~=~ \sum_{m,n}\,z^{-\frac{3+s}{2}-m}\bz^{-\frac{3-s}{2}-n}\,W^s_{m,n}~.
    \label{equ:Ws-mode-exp}
\end{equation}
The $Lw_{1+\infty}$ algebra can then be recast as
\begin{equation}
    \begin{split}
        \Big[W_s(z,\bz), W_{s'}(w,\bw)\Big] ~=&~ \frac{i}{2}\,\Big[ (s+1)\pa_w\delta^{(2)}(z-w)\,W_{s+s'-1}(z,\bz) \\
        &\qquad ~-~ (s'+1)\pa_z\delta^{(2)}(z-w)\,W_{s+s'-1}(w,\bw) \Big]~~,
    \end{split}
    \label{equ:w1+infty-4D-1}
\end{equation}
or, equivalently (using (\ref{equ:deltafun-deriv}) and (\ref{equ:delfuc-manipulation})
\begin{equation}
    \begin{split}
        \Big[W_s(z,\bz), W_{s'}(w,\bw)\Big]  ~=&~ \frac{i}{2}\,\Big[ (s+s'+2)\pa_w\delta^{(2)}(z-w)\,W_{s+s'-1}(w,\bw) \\
        &\qquad~+~ (s+1)\delta^{(2)}(z-w)\,\pa_wW_{s+s'-1}(w,\bw)\Big]~~.
    \end{split}
       \label{equ:w1+infty-4D}
\end{equation}
After plugging in the mode expansion (\ref{equ:w1+infty-4D-1}), one can see that this reduces to the Loop algebra
\begin{equation}
    \Big[ W^s_{m,n}, W^{s'}_{p,q} \Big] ~=~ \frac{i}{2}\,\Big[ m(s'+1) - p(s+1) \Big]\,W_{m+p,n+q}^{s+s'-1} ~~,
\end{equation}
which matches the barred version of~\eqref{eq:wloop} after the mode redefinition $W_m^s(\bz)=w^{\frac{s+3}{2}}_{\frac{s+1}{2}+m}(\bz)$.

Now, building off their previous work~\cite{Freidel:2021qpz,Freidel:2021dfs}, Freidel et al.~\cite{Freidel:2021ytz} were able to reorganize the asymptotic vacuum Einstein equations into a recursion relation among a tower of spin-$s$ modes $q_s(u,z,\bz)$  where the first entry is the news
\be
q_{-2}(u,z,\bz)=\frac{1}{2}\pa_uN^{zz}~~,
\ee
and for $s\le3$ the recursion takes the form
\begin{equation}
    q_s ~=~ \pa_u^{-1}\,D_z\,q_{s-1} ~+~ \frac{(s+1)}{2}\,\pa_u^{-1}\Big(C\,q_{s-2}\Big)~~,
    \label{equ:qs-recursion}
\end{equation}
where the inverse $\pa_u$ in~\eqref{equ:qs-recursion} is defined as
\begin{equation}
    \pa_u^{-n}\,f(u) ~:=~ \int_{+\infty}^u\,du_1\,\cdots\,\int_{+\infty}^{u_{n-1}}\,du_n\,f(u_n) ~~.
\end{equation}
To match the vacuum Einstein equations, this recursion needs to be modified by terms quadratic in $\bar C$ starting at $s=4$, thought to be related to the mixed helicity OPE. Nevertheless, the authors of~\cite{Freidel:2021ytz} showed that solutions to~\eqref{equ:qs-recursion}  reproduce the $Lw_{1+\infty}$ symmetry.

Given~\eqref{equ:qs-recursion} and~\eqref{equ:q1s} one can see that terms with larger $s$ will involve more gravitons.  To solve this recursion relation, they expanded the charges based on the number of oscillators 
\be
q_s=\sum_{k=1}^{\max(2,s+1)} q^k_s~~,
\ee after which (\ref{equ:qs-recursion}) reduces to
\begin{equation}
    q^k_s ~=~ \pa_u^{-1}\,D_z\,q^k_{s-1} ~+~ \frac{(s+1)}{2}\,\pa_u^{-1}\Big(C\,q^{k-1}_{s-2}\Big)~~.
    \label{equ:qks-recursion}
\end{equation}
The solutions for the first two oscillator-orders are
\begin{align}
q_{s}^1(u,z,\bz) ~=&~ \frac{1}{2}\,(\pa_u^{-1}\,D_z)^{s+2}\,\pa_u\,N^{zz}(u,z,\bz)~~, \label{equ:q1s}\\
    q_{s}^2(u,z,\bz) ~=&~ \frac{1}{4}\,\sum_{l=0}^s\,(l+1)\,\pa_u^{-1}\,\Big( \pa_u^{-1}\,D_z\Big)^{s-l}\,\Big[C\Big( \pa_u^{-1}\,D_z\Big)^l\pa_u^2\bar{C}\Big]~~.
    \label{equ:q2s}
\end{align}
The $W_s$ symmetry generators are then defined to be the $q_s$ evaluated at the corner. To actually take the $u\to-\infty$ limit, they need to introduce a renormalization prescription from \cite{Freidel:2021dfs} 
\begin{equation}
    \hat{q}_s(u,z,\bz) ~=~ \sum_{n=0}^s\,\frac{(-u)^{s-n}}{(s-n)!}\,D_z^{s-n}\,q_n(u,z,\bz)~,
    \label{equ:qs-renormalization}
\end{equation}
after which the higher spin charge aspects are defined as 
\begin{equation}
    W_s(z,\bz) ~=~ \lim_{u\to-\infty}\,\hat{q}_s(u,z,\bz)~.
    \label{equ:qscharge}
\end{equation}
One can then check that the linear truncation of the Poisson bracket
\begin{equation}\scalemath{.95}{
   \begin{aligned}
        \{W_s(z,\bz),&\,W_{s'}(w,\bw)\}^1 ~=~ \{W^1_s(z,\bz),\,W^2_{s'}(w,\bw)\} ~+~ \{W^2_s(z,\bz),\,W^1_{s'}(w,\bw)\}\\
        ~=&~ \frac{\kappa^2}{8}\,\Big[ (s+1)D_w\delta^{(2)}(z-w)\,W^1_{s+s'-1}(z,\bz) ~-~ (s'+1)D_z\delta^{(2)}(z-w)\,W^1_{s+s'-1}(w,\bw) \Big] ~~
   \end{aligned}}
\end{equation}
matches the $w_{1+\infty}$ algebra observed from the radial commutation relations. While the cubic graviton vertex contributed to the splitting function in the OPE derivation, here it is the Poisson brackets of the two-oscillator $k=2$ terms with the conformally soft single graviton modes at $k=1$ that reproduce the algebra. This should be reminiscent of the original soft/hard splitting of the gravitational asymptotic symmetry charges~\cite{He:2014laa,Kapec:2014opa}, where the hard gravitational charge induces the homogeneous transformations of the gravitational field. It begs the question: can we add matter to this construction?

\subsection{\texorpdfstring{Conformally Soft Multi-Particle Operators and $w_{1+\infty}$}{Conformally Soft Multi-Particle Operators and w1+infty}}\label{sec:algresult}

From our discussion in section~\ref{sec:LRcorner}, we saw that in the cases where the algebra of conformal collider operators could be matched to an asymptotic symmetry, they were simply the matter charge acting on the out state. Namely when we couple to gravity we get both a soft and hard splitting
\be
Q^{{\cal I}^+}=Q_S^{{\cal I}^+}+Q_H^{{\cal I}^+},~~~Q^{{\cal I}^+}_H=Q^{{\cal I}^+}_{H,grav}+Q^{{\cal I}^+}_{H,matter}~~,
\label{eq:q_split}
\ee
and an additional splitting of the hard charge into a gravitational and a non-gravitational part. In particular, for the leading soft graviton theorem, the hard matter charge is the ANEC operator in the matter theory, while the full hard charge is the shear-inclusive ANEC. Meanwhile, the soft charge is linear in the matter field, as required to generate the inhomogeneous shift expected of supertranslations. 

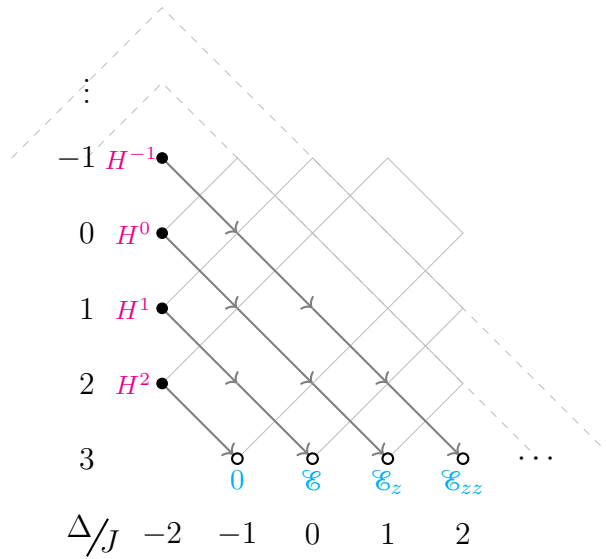
\begin{figure}[htp!]
 \centering
 \begin{tikzpicture}[scale=1]
\definecolor{darkgreen}{rgb}{.0, 0.5, .1};
\draw[lightgray] (0,2) -- (3,5) -- (4,4) -- (1,1);
\draw[lightgray] (0,3) -- (2,5) -- (4,3) -- (2,1);
\draw[lightgray] (0,4) -- (1,5) -- (4,2) -- (3,1);
\draw[lightgray,dashed] (-1,5) -- (0,6) --  (5,1);
\draw[lightgray,dashed] (-2,5) -- (0,7) --  (6,1);
\draw[->,gray,thick] (0,2) -- (1-.05,1+.05);
\draw[->,gray,thick] (0,3) -- (1,2);
\draw[->,gray,thick]  (1,2)--(2-.05,1+.05);
\draw[->,gray,thick] (0,4) -- (1,3);
\draw[->,gray,thick] (1,3) -- (2,2);
\draw[->,gray,thick] (2,2) -- (3-.05,1+.05);
\draw[->,gray,thick] (0,5) -- (1,4);
\draw[->,gray,thick] (1,4) -- (2,3);
\draw[->,gray,thick] (2,3) -- (3,2);
\draw[->,gray,thick] (3,2) -- (4-.05,1+.05);
\node at (3,0) {$1$};
\node at (4,0) {$2$};
\draw[thick,fill=white] (1,1) circle (2pt) node[below]{\color{cyan}\small $0$} ;
\draw[thick,fill=white] (2,1) circle (2pt) node[below]{\color{cyan}\small $\cal E$} ;
\draw[thick,fill=white] (3,1) circle (2pt) node[below]{\color{cyan}\small $\mathcal{E}_z$} ;
\node at (5,1) {$\cdots$};
\draw[thick,fill=white] (4,1) circle (2pt) node[below]{\color{cyan}\small $\mathcal{E}_{zz}$} ;
\node at (-.7,2.9-3)  {$J$};
\node at (-1.1,3.1-3) {$\tiny{\Delta}$};
\draw[thick] (-1,-.3) -- (-1+.28,.2);
\filldraw[black] (0,2) circle (2pt) node[left]{\color{magenta}\footnotesize $H^2$} ;
\filldraw[black] (0,3) circle (2pt) node[left]{\color{magenta}\footnotesize $H^1$} ;
\filldraw[black] (0,4) circle (2pt) node[left]{\color{magenta}\footnotesize $H^0$} ;
 \filldraw[black] (0,5) circle (2pt) node[left,xshift=.2em]{\color{magenta}\footnotesize $H^{-1}$} ;
\node at (-1,1) {$3$};
\node at (-1,2) {$2$};
\node at (-1,3) {$1$};
\node at (-1,4) {$0$};
\node at (-1,5) {$-1$~~~};
\node at (-1,6) {$\vdots$};
\node at (0,0) {$-2$};
\node at (1,0) {$-1$};
\node at (2,0) {$0$};
\end{tikzpicture}
\caption{Celestial conformal dimensions and spins of the graviton modes giving rise to a $w_{1+\infty}$ symmetry (magenta), and modes in the matter sector we expect them to couple to (cyan). The spectrum of these operators are related by 2D light transform.  For reference, the `celestial diamonds' are superimposed in grey.
\label{fig:diamonds}}
 \end{figure}

Despite it being quadratic in the fields, the fact that the ANEC operator is a celestial primary then fits nicely into the existing celestial CFT literature: this operator has the same conformal dimensions as the $\Delta=1$ graviton's level-2 primary descendant, and the Ward identity is simply a null state in the CCFT that follows from the asymptotic Einstein equations and antipodal matching. Phrased in terms of the `current' $P_\bz=\frac{1}{8\pi G}D^\bz N_{\bz\bz}$ used in~\cite{He:2014laa,Himwich:2020rro} (up to a convenient choice of normalization), equation~\eqref{eq:ward} becomes
\be\label{Pnull}
\p_z (P_\bz^+-P_\bz^-)={\cal E}^+-{\cal E}^-~,
\ee 
where the $\pm$ superscripts distinguish operators at $\mathcal{I}^\pm$. Namely, this hard charge sources the (generalized) current.  This is nicely captured by the celestial diamond framework~\cite{Pasterski:2021fjn,Pasterski:2021dqe}.   As illustrated in Figure~\ref{fig:diamonds}, this story nominally extends to a tower of modes with dimension $\Delta=3$ that can appear in the spectrum of primary descendants of the conformally soft graviton modes. The $w_{1+\infty}$ generators are actually more cleanly written in terms of the light transforms of these graviton modes (see e.g.~\cite{Himwich:2021dau}).  These light transformed gravitons have precisely the same spectrum as the expected matter sources. 

It is important to note that despite the fact that the ANEC operator vanishes away from other particle insertions in an $\cal S$-matrix element with a finite number of particles in the out state,\footnote{Hence the left hand side of~\eqref{Pnull} is often set to zero as an operator equation / treated as the null state in CCFT.} guided by symmetries, we can still study the generalized detector operators relevant to the $w_{1+\infty}$ extension using the phase space approach outlined in section~\ref{sec:winftyalg}.  Moreover, we definitely don't want to assume $\cal E$ will vanish almost everywhere for generic out states in a collider experiment when we have a finite angular resolution for our calorimeters.

While~\cite{Freidel:2021ytz} only considered the vacuum Einstein equations, the earlier work~\cite{Freidel:2021qpz} arranged the full matter coupled equations in a similar manner, by definite spin. The cyan labels in Figure~\ref{fig:diamonds} reflect their notation. The $u$-zero modes of the spin-0 and spin-1 components reproduce the BMS subalgebra, whose matter charges were recast in a conformal collider context by Cordova and Shao~\cite{Cordova:2018ygx}. The structure of the $w_{1+\infty}$ is such that only these $W_{s\le1}$ form a closed subalgebra. Once you have the $(\Delta,J)=(3,2)$ mode, you can generate the rest of the tower. We thus expect the construction `has to work' once we are confident in the sub-subleading soft graviton theorem~\cite{Cachazo:2014fwa,Campiglia:2016jdj,Campiglia:2016efb,Laddha:2017ygw,Freidel:2021dfs}. Inspired by~\eqref{eq:q_split}
we might expect a splitting of the form
\be
W_s=W_{s,grav}+W_{s,matter}~~,
\label{eq:Wsplit}
\ee
where, as in the $s=0,1$ subalgebra, the only single oscillator $k=1$ piece is in $W_{s,grav}$, while the matter charge starts at $k=2$. However, we will see that we need higher oscillator terms and, in particular, mixed graviton/matter terms starting at $k=3$ to get the algebra to close when we go to $s\ge 2$.

\paragraph{Adding Matter} Say we wanted to reproduce the $Lw_{1+\infty}$ algebra starting from the matter phase space. One advantage of the analysis in \cite{Freidel:2021ytz} is that they don't need to restrict to the single helicity sector and their charge brackets don't rely on the helicity carried by the fields or oscillators. In other words, starting from (\ref{equ:qks-recursion}-\ref{equ:qscharge}), one can replace all the $C$ and $\bar{C}$ by matter fields and the resulting charge bracket should remain the same once the matter field modes admit the same commutation relations. With that in mind, let's consider a free complex scalar theory where the large-$r$ expansion is as follows\footnote{In what follows we will suppress the superscript $(0)$ and $\phi(u,z,\bz)=\phi^{(0)}(u,z,\bz)$. Also, we skip the step of taking $\lim_{r\to\infty}r^2$ and present the final results directly.}.
\begin{equation}
    \phi(u,r,z,\bz) ~=~ \frac{\phi^{(0)}(u,z,\bz)}{r} ~+~ {\cal O}\left(\frac{1}{r^2}\right) ~~,~~ \bar{\phi}(u,r,z,\bz) ~=~ \frac{\bar{\phi}^{(0)}(u,z,\bz)}{r} ~+~ {\cal O}\left(\frac{1}{r^2}\right)~~. 
\end{equation}
The mode expansion of $\phi^{(0)}(u,z,\bz)$ can be obtained by simply replacing the oscillators $a_{\pm}(a_{\pm}^{\dagger})$ in (\ref{equ:mode-exp-C}) by $a(a^{\dagger})$ and $b(b^{\dagger})$ with the same commutation relation (\ref{eq:adaggera})
\begin{align}
    \phi^{(0)}(u,z,\bz) ~=&~ \frac{i}{(2\pi)^2}\,\int_0^{\infty}d\omega\,\Big[ a^{\dagger}(\omega,z,\bz)e^{i\omega u} - b(\omega,z,\bz)e^{-i\omega u}\Big] ~~,\\
    \bar{\phi}^{(0)}(u,z,\bz) ~=&~ \frac{i}{(2\pi)^2}\,\int_0^{\infty}d\omega\,\Big[ b^{\dagger}(\omega,z,\bz)e^{i\omega u} - a(\omega,z,\bz)e^{-i\omega u}\Big] ~~.
\end{align}
Then, the quadratic matter operator becomes
\begin{equation}
    W^2_{s,matter}(z,\bz) 
        ~=~ -\frac{1}{4}\,\sum_{l=0}^s\,(-1)^{s-l}\,\frac{(l+1)}{(s-l)!}\,\int_{-\infty}^{+\infty}\,du\,u^{s-l}\,\pa^{s-l}_z\,\Big[\, \phi(u,z,\bz)\,\pa_z^l\pa^{2-l}_u\bar{\phi}(u,z,\bz)\,\Big]~~,
        \label{equ:W2s-matter}
\end{equation}
which has $(\Delta,J) = (3,s)$. 
 Note that both ${\cal L}_n$-type and ${\cal E}_J$-type operators occur in this expression. Specifically, the $l=s$ term is of ${\cal E}_J$-type and the $l=0$ term contains the ${\cal L}_n$-type. In particular, for the $s=0,1$ modes we have
\begin{equation}
    \begin{split}
        W_{0,matter}^2(z,\bz) ~=&~ -\frac{1}{4}\,{\cal E}_2(z,\bz)~, \\
        W_{1,matter}^2(z,\bz) ~=&~ \frac{1}{4}\,D_z\,{\cal L}_{-1}(z,\bz) ~-~ \frac{1}{2}\,{\cal N}_z(z,\bz) ~-~ \frac{i}{4}\,D_z\,{\cal E}_1(z,\bz)~,  \\
    \end{split}
\end{equation}
which form a closed subalgebra. This agrees with the BMS subalgebra identified in~\cite{Cordova:2018ygx} up to an additional ${\cal E}_1$ term that doesn't modify the commutation relations.  

As a further check of the general-$s$ modes, we can look at the graviton-scalar OPE. Recall that, for the gravitational case, the algebra of the quadratic graviton modes with the single particle modes reproduced the result extracted from the OPE. If this were to carry over to our ansatz~\eqref{eq:Wsplit}, then the relevant term to match is $[W_s,\phi_{\Delta}]^1$, which would come from $W_{s,matter}^2$.
Plugging in~\eqref{equ:W2s-matter} we find
\begin{equation}
\scalemath{.93}{
    \begin{aligned}
        \Big[W^2_{s}(z,\bz),\phi_{\Delta}(w,\bw)\Big]^1 ~=&~ \Big[W^2_{s,matter}(z,\bz), \Gamma(\Delta)\int\,du\,u^{-\Delta}\,\phi(u',w,w)\Big]^1\\
        ~=&~ -\frac{i}{2}\,\sum_{l=0}^s\,(-1)^{s+l}\,\frac{(l+1)(\Delta)_{s-l}}{(s-l)!}\,\pa_z^{s-l}\delta^{(2)}(z-w)\,\pa_w^l\phi_{\Delta+1-s}(w,\bw)~.
    \end{aligned}}
\end{equation}
Using (\ref{equ:deltafun-mode-exp}) and (\ref{equ:Ws-mode-exp}) we can then extract
\begin{equation}
    \begin{split}
        \Big[ W^{2,s}_{m,n},\phi_{\Delta}(w,\bw)\Big] ~=&~ -\frac{i}{2}\,\sum_{l=0}^s\,(-1)^{l}\,\frac{(s-l+1)(\Delta)_{l}}{l!}\,\frac{\Gamma(l-\frac{s+3}{2}-m+1)}{\Gamma(1-\frac{s+3}{2}-m)}\\
        &\qquad\qquad w^{\frac{1+s}{2}+m-l}\bw^{\frac{1-s}{2}+n}\,\pa_w^{s-l}\phi_{\Delta+1-s}(w,\bw)~.
    \end{split}
\end{equation}
Applying the Gamma function identity (\ref{equ:GammafuncID}), we have
\begin{equation}
    (-1)^{l}\frac{\Gamma(l-\frac{s+3}{2}-m+1)}{l!\Gamma(1-\frac{s+3}{2}-m)} ~=~ \frac{\Gamma(\frac{s+3}{2}+m)}{l!\Gamma(\frac{s+3}{2}+m-l)} ~=~ \begin{pmatrix}
    \frac{s+3}{2}+m-1\\
    l
    \end{pmatrix}~,
\end{equation}
so that
\begin{equation}\scalemath{.92}{
    \badat{3}
        \Big[ W^{2,s}_{m,n},\phi_{\Delta,J}(w,\bw)\Big] =& -\frac{i}{2}\,\sum_{l=0}^s\,\begin{pmatrix}
    \frac{s+3}{2}+m-1\\
    l
    \end{pmatrix}(s-l+1)(\Delta)_{l}\, w^{\frac{1+s}{2}+m-l}\bw^{\frac{1-s}{2}+n}\,\pa_w^{s-l}\phi_{\Delta+1-s}(w,\bw)~.
    \eadat}
    \label{equ:W2mn-phi-mode}
\end{equation}
A general analysis of celestial currents constructed from conformally soft gravitons in tree-level minimally-coupled gravitational theories was carried out in~\cite{Himwich:2021dau}. There they derived the $w_{1+\infty}$ actions on arbitrary massless primaries. Plugging in $n=\frac{1}{2}(s-1)$ we see that (\ref{equ:W2mn-phi-mode}) matches the complex conjugate of their $[\hat{w}^{q}_m,{\cal O}_{\Delta,J}]$ action for $q=\frac{1}{2}(s+3)$ and $J=0$. As a nontrivial check, one can show that the $[W_{s}^2,\phi_{\Delta,J}]^1$ commutator continues to match the OPE-derived expression when we promote our complex scalar to a massless spinning field in all of our expressions.

However, the commutator between $W^2_{s}$'s doesn't close for $s\ge 2$. This is not an issue for the gravitational modes precisely because of a cancellation coming from the linear and cubic terms. In principle, one could follow the same substitution that worked at quadratic order and consider the linear and cubic modes (here we show the explicit expression for $s=2$)
\begin{equation}
    \begin{split}
        W_{s,matter}^1(z,\bz) ~\overset{?}{=}&~ \frac{1}{2}\frac{(-1)^{s+1}}{s!}\,\int_{-\infty}^{+\infty}du\,u^s\, \pa_z^{s+2}\,\pa_u\,\bar{\phi}(u,z,\bz)~,\\
        W_{2,matter}^3(z,\bz) ~\overset{?}{=}&~ -\frac{3}{2}\,\frac{1}{4}\,\int_{-\infty}^{+\infty}du\,\Big[ \phi(u,z,\bz)\,\int^{u}_{+\infty}du'\,\Big( \phi(u',z,\bz)\,\pa_{u'}^2\,\bar{\phi}(u',z,\bz) \Big) \Big]~.
    \end{split}
\end{equation}
Then the algebra again closes and one reproduces the $w_{1+\infty}$ through quadratic order. However, while this substitution of the scalar mode operators in place of the graviton ones formally works at the level of the canonical commutation relations, $W_s^1$ now has $(\Delta,J) = (3,s+2)$ and $W_s^3$ has $(\Delta,J) = (3,s-2)$, i.e. the $W_s$ generator is a linear combination of operators with different spins. While in the end you are just rearranging symmetries of the free theory in an awkward way, given the global symmetries that survive in the perturbative limit, one would expect that the generator $W_s$ should have a definite spin. Moreover, there is no reason to expect the cubic scalar interaction in a given theory to obey any universal symmetry structure, as our interpretation of $[W_s,W_s']^1$ above would imply.

We mention this because one would seem to encounter this kind of construction if one wanted a purely matter representation. We wanted to start with just the quadratic terms, mimicking the ANEC and its superrotation analog, but end up needing to add single particle (and cubic) terms as well in order to get closure of the quadratic modes. This runs counter to our assumptions about the $k=1$ modes. However, if we stick to the notion that these higher oscillator modes are supposed to complete the conformally soft graviton modes so that the phase space brackets reproduce the OPEs we get from the collinear splitting functions and, from the OPEs, we indeed want to have only graviton contributions to $k=1$, then we can cancel the problematic terms in the $k=2$ matter modes by introducing mixed graviton/matter modes at $k=3$.

In summary, the $w_{1+\infty}$ generators can be organized according to the number of oscillators $k$
\begin{equation}
    W_s(z,\bz) ~=~ \sum_{k=1}^{\max(2,s+1)} W_s^k(z,\bz) ~,
\end{equation}
where at each level, $W_s^k(z,\bz)$ has conformal weight 3 and spin $s$. 
We will start by taking the full tower of generators from gravity into account, adding the quadratic matter operator, and then introduce conformally soft higher-particle operators by coupling soft gravitons with matter fields so that the algebra closes order by order. The rest of this section will be devoted to presenting the expressions of operators up to the cubic level, explicitly showing that they realize the $w_{1+\infty}$ algebra (\ref{equ:w1+infty-4D}) up to quadratic order, and connecting this construction to our results from section~\ref{sec:LRcorner}.

\paragraph{Linear Operators} At the linear order, the $W_s^1$ generator is simply the $q_s^1$ from gravity. Namely, 
\begin{equation}
    W_s^1(z,\bz) ~=~ \frac{1}{2}\frac{(-1)^{s+1}}{s!}\,\int_{-\infty}^{+\infty}du\,u^s\, D_z^{s+2}\,\pa_u\,\bar{C}(u,z,\bz)~.
\end{equation}
After plugging in the mode expansion (\ref{equ:mode-exp-C}), we can write this as
\begin{equation}
       W_s^{1,-}(z,\bz) ~=~ \frac{1}{2}\frac{(-1)^{s+2}}{s!}\,\frac{i^s}{(2\pi)}\,\int_0^{\infty}d\omega\,\omega\,\pa_{\omega}^s\delta(\omega)\,D_z^{s+2}\,a_-(\omega,z,\bz)~,
       \label{equ:W1s-mode}
\end{equation}
where the commutation between $a_-$ and $a^{\dagger}_-$ is normalized as in~\eqref{eq:adaggera}. Here we've suppressed an analogous $W_s^{1,+}$ term involving $a_+^\dagger$. This will simplify our expressions in what follows since the algebra for these terms will take the same form.

\paragraph{Quadratic Operators} At the quadratic level, we consider two types of operators, coming from the gravitational and matter sectors, respectively 
\begin{equation}\scalemath{.97}{%
    \begin{aligned}
        W^2_{s,grav}(z,\bz) 
        ~=&~ -\frac{1}{4}\,\sum_{l=0}^s\,(-1)^{s-l}\,\frac{(l+1)}{(s-l)!}\,\int_{-\infty}^{+\infty}\,du\,u^{s-l}\,D^{s-l}_z\,\Big[\, C(u,z,\bz)\,D_z^l\pa^{2-l}_u\bar{C}(u,z,\bz)\,\Big]~,\\
                W^2_{s,matter}(z,\bz) 
        ~=&~ -\frac{1}{4}\,\sum_{l=0}^s\,(-1)^{s-l}\,\frac{(l+1)}{(s-l)!}\,\int_{-\infty}^{+\infty}\,du\,u^{s-l}\,D^{s-l}_z\,\Big[\, \phi(u,z,\bz)\,D_z^l\pa^{2-l}_u\bar{\phi}(u,z,\bz)\,\Big]~.
    \end{aligned}}%
\end{equation}
After plugging in the mode expansions, only the terms with one creation and one annihilation operator survive the $u$-integrals 
\begin{equation}
\scalemath{.97}{
    \begin{aligned}
     W^{2,-}_{s,grav}(z,\bz) ~=&~ \frac{1}{4(2\pi)^3}\,\sum_{l=0}^s\,(-1)^{s-l}(-i)^{s+2-2l}\,\frac{(l+1)}{(s-l)!}\\
        & \int_{0}^{+\infty}\,d\omega\,\int_{0}^{+\infty}\,d\omega'\,\omega'^{2-l}\pa_{\omega}^{s-l}\,\delta(\omega-\omega') D^{s-l}_z\,\Big[\, a_-^\dagger(\omega,z,\bz)\,D_z^l a_-(\omega',z,\bz)\,\Big]~~,\\
        W^{2,-}_{s,matter}(z,\bz) ~=&~ \frac{1}{4(2\pi)^3}\,\sum_{l=0}^s\,(-1)^{s-l}(-i)^{s+2-2l}\,\frac{(l+1)}{(s-l)!}\\
        & \int_{0}^{+\infty}\,d\omega\,\int_{0}^{+\infty}\,d\omega'\,\omega'^{2-l}\pa_{\omega}^{s-l}\,\delta(\omega-\omega') D^{s-l}_z\,\Big[\, a^\dagger(\omega,z,\bz)\,D_z^l a(\omega',z,\bz)\,\Big]~~
    \end{aligned}}
    \label{equ:mode-W2s}
\end{equation}
where, we are similarly suppressing terms involving the $b$ oscillator which appear in $W^{2,+}_{s,matter}$.

\paragraph{Cubic Operators} When $s\ge 2$, cubic operators show up. From (\ref{equ:qks-recursion}) and (\ref{equ:qs-renormalization}) we have the following cubic charges from gravity. 
\begin{equation}
    W_{s,grav}^3(z,\bz) ~=~ \lim_{u\to -\infty}\,\sum_{n=0}^s\,\frac{(-u)^{s-n}}{(s-n)!}\,D^{s-n}_z\,q_s^3(u,z,\bz)~,
\end{equation}
where 
\begin{equation}
    q_s^3(u,z,\bz) ~=~  D_z\pa_u^{-1}q_{s-1}^3 + \frac{s+1}{2}\,\pa_u^{-1}(C\,q_{s-2}^2) ~,
\end{equation}
and the expression of $q_{s}^2(u,z,\bz)$ is presented in (\ref{equ:q2s}). 
Particularly, when $s=2$, we have
\begin{equation}
    W_{2,grav}^3(z,\bz) ~=~ \frac{3}{2}\,\frac{1}{4}\,\pa_{u}^{-1}\Big[ C\,\pa_u^{-1}\Big( C\,\pa_u^2\,\bar{C} \Big) \Big] ~.
    \label{equ:W23-grav-field}
\end{equation}
Coupled to the matter sector, we introduce the following matter cubic operator at $s=2$,
\begin{equation}
    \begin{aligned}
   W_{2,matter}^3(z,\bz) ~=&~ \frac{3}{2}\,\frac{1}{4}\,\pa_{u}^{-1}\Big[ C\,\pa_u^{-1}\Big( \phi\,\pa_u^2\,\bar{\phi} \Big) ~+~ \phi\,\pa_u^{-1}\Big( C\,\pa_u^2\,\bar{\phi} \Big) \Big] ~,
    \end{aligned}
    \label{equ:W23-matter-field}
\end{equation}
which, after plugging in the mode expansions, reduces to
\begin{equation}
    \begin{aligned}
   W_{2,matter}^{3,-}(z,\bz) 
   ~=&~ \frac{3}{2}\frac{1}{4}\,\frac{i}{(2\pi)^6}\,\int_{-\infty}^{+\infty}\,du\,\int_0^{\infty}d\omega_1\,a_-^{\dagger}(\omega_1,z,\bz)\,e^{i\omega_1u}\int_{+\infty}^{u}\,du'\\
       &\qquad\qquad \int_0^{\infty}d\omega_2\,a^{\dagger}(\omega_2,z,\bz)\, \int_0^{\infty}d\omega_3\,\omega_3^2\,e^{i(\omega_2-\omega_3)u'}\,a(\omega_3,z,\bz)\\
       &~+~ \frac{3}{2}\frac{1}{4}\,\frac{i}{(2\pi)^6}\,\int_{-\infty}^{+\infty}\,du\,\int_0^{\infty}d\omega_1\,a^{\dagger}(\omega_1,z,\bz)\,e^{i\omega_1u}\int_{+\infty}^{u}\,du'\\
       &\qquad\qquad \int_0^{\infty}d\omega_2\,a_-^{\dagger}(\omega_2,z,\bz)\, \int_0^{\infty}d\omega_3\,\omega_3^2\,e^{i(\omega_2-\omega_3)u'}\,a(\omega_3,z,\bz)~.\\
    \end{aligned}
    \label{equ:W23-mode}
\end{equation}
Since we are interested in checking the matter algebra in what follows we omit the mode expansion for the cubic graviton terms here. The cubic term is all that we will need to check the algebra to quadratic order, meanwhile the $s\le2$ terms are enough to generate the rest of the algebra. 

\paragraph{Checking the Algebra} Now we are ready to check the algebra generated by $W_s(z,\bz)$ operators. Clearly, since 
\begin{equation}
   \Big[ W_s^1(z,\bz),\, W^2_{s',matter}(w,\bw) \Big] ~=~ 0~,
\end{equation}
the linear truncation reduces to the derivation in \cite{Freidel:2021ytz} and we have
\begin{equation}
    \begin{split}
        \Big[ W_s(z,\bz),\,W_{s'}(w,\bw) \Big]^1 ~=&~  \Big[ W^1_s(z,\bz),\,W^2_{s',grav}(w,\bw) \Big] ~+~  \Big[ W^2_{s,grav}(z,\bz),\,W^1_{s'}(w,\bw) \Big]\\
        ~=&~ \frac{i}{2}\,\Big[ (s+s'+2)\,D_w\delta^{(2)}(z-w)\,W_{s+s'-1}^1(w,\bw) \\
        &\qquad ~+~ (s+1)\,\delta^{(2)}(z-w)\,D_w\,W_{s+s'-1}^1(w,\bw) \Big]~.
    \end{split}
\end{equation}
We will now show that the $w_{1+\infty}$ continues to hold at quadratic order. As discussed in~\cite{Guevara:2021abz}, the entire tower of symmetries is generated by $W_{s\le 2}$. The $W_{s\le1}$ modes form a closed subalgebra while the rest of higher spin modes are generated from the commutators involving $W_2$. Therefore, we will focus on the commutation relations between $W_0,W_1,$ and $W_2$. Generally speaking, the quadratic level receives the following contributions
\begin{equation}
    \begin{split}
        \Big[ W_s(z,\bz),\,W_{s'}(w,\bw) \Big]^2 ~=&~  \Big[ W_s(z,\bz),\,W_{s'}(w,\bw) \Big]^2_{grav} ~+~ \Big[ W_s(z,\bz),\,W_{s'}(w,\bw) \Big]^2_{matter}~~,
    \end{split}
\end{equation}
where
\begin{equation}
    \badat{3}
        &\Big[ W_s(z,\bz),\,W_{s'}(w,\bw) \Big]^2_{grav/matter} ~=~  \Big[ W^2_{s,grav/matter}(z,\bz),\,W^2_{s',grav/matter}(w,\bw) \Big]\\
       & ~~~~~~~~~~~~~~~~~+~ \Big[ W^1_s(z,\bz),\,W^3_{s',grav/matter}(w,\bw) \Big] 
         ~+~ \Big[ W^3_{s,grav/matter}(z,\bz),\,W^1_{s'}(w,\bw) \Big]~.
\eadat\label{eq:w2}
\end{equation}
The terms in the second line distinguish the structure of the gravity and matter cancellations. One can remove the matter sector and still cancel the offending terms in $[W_s^2,W_{s'}^2]_{grav}$ to restore closure. The same is not true if we try to decouple gravity from our theory. In that case, only the $W_{s\le1,matter}$ subalgebra closes, as we will now check.

\paragraph{BMS Subalgebra}
For $s,s'=0,1$ there are no contributions from cubic operators and the commutator becomes
\begin{equation}\resizebox{0.9\textwidth}{!}{$%
    \Big[ W_s(z,\bz),\,W_{s'}(w,\bw) \Big]^2 ~=~ \Big[ W^2_{s,grav}(z,\bz),\,W^2_{s',grav}(w,\bw) \Big]~+~ \Big[ W^2_{s,matter}(z,\bz),\,W^2_{s',matter}(w,\bw) \Big]~. $}%
\end{equation}
Direct calculations starting from (\ref{equ:mode-W2s}) yield 
\begin{align}
    &\Big[ W_0(z,\bz),\,W_{0}(w,\bw) \Big]^2_{matter/grav} ~=~ 0~~,\\
    \begin{split}
        \Big[ W_0(z,\bz),\,W_{1}(w,\bw) \Big]^2_{matter/grav} ~=&~  \frac{i}{2}\,\Big[ 3\,D_{w}\delta^{(2)}(z-w)\,W^2_{0,matter/grav}(w) \\
    &\qquad + \delta^{(2)}(z-w)\,D_w\,W^2_{0,matter/grav}(w) \Big]~~,\\
    \end{split}\\
    \begin{split}
        \Big[ W_1(z,\bz),\,W_{1}(w,\bw) \Big]^2_{matter/grav} ~=&~ \frac{i}{2}\,\Big[ 4\,D_{w}\delta^{(2)}(z-w)\,W^2_{1,matter/grav}(w) \\
    &\qquad + 2\,\delta^{(2)}(z-w)\,D_w\,W^2_{1,matter/grav}(w) \Big]~~,\\
    \end{split}
\end{align}
which exactly matches with $w_{1+\infty}$ (\ref{equ:w1+infty-4D}). Moreover, note that the gravity sector and matter sector decouple in the subalgebra, matching the splitting discussed below~\eqref{eq:q_split}. If one turns off gravity the algebra still holds, a feature that underpins the computations in~\cite{Cordova:2018ygx}.

\paragraph{Adding $W_2$} For the $s=2$ mode we now expect contributions from the cubic operator $W_2^3$. We want to show that the commutator between $W_s^1$ and $W_{2,matter/grav}^3$ makes our algebra close. As seen from (\ref{equ:mode-W2s}) and (\ref{equ:W23-grav-field}-\ref{equ:W23-matter-field}), the modes $W^2_{2,matter}$, $W^3_{2,matter}$ and $W^2_{2,grav}$, $W^3_{2,grav}$ have a similar structure, so we will only focus on the matter case here. Since the $[W_2,W_2]$ commutators generate the $W_{s>2}$ modes, we will only need to explicitly check $[W_0,W_2]$ and $[W_1,W_2]$.

\paragraph{$\mathbf{\succ}~s=0, s'=2:$}
Direct computation gives
\begin{equation}\resizebox{0.9\textwidth}{!}{$%
    \begin{aligned}
            \Big[ W^{2,-}_{0,matter}(z), W^{2,-}_{2,matter}(w)\Big] ~=&~ \frac{i}{2}\,\Big[ 4\,\pa_{w}\delta^{(2)}(z-w)\,W^{2,-}_{1,matter}(w) + \delta^{(2)}(z-w)\,\pa_w\,W^{2,-}_{1,matter}(w) \Big] \\
    &~+~ \frac{1}{4}\frac{3}{2}\,\pa^2_{w}\delta^{(2)}(z-w)\,\frac{1}{(2\pi)^3}\,\Bigg\{ \int_0^{\infty}d\omega\,\omega\,a^{\dagger}(\omega,w)a(\omega,w)\\
    &\qquad\quad ~+~ \int_0^{\infty}d\omega_2\int_0^{\infty}d\omega_3\omega_3^2\pa_{\omega_2}\delta(\omega_2-\omega_3)\,a^{\dagger}(\omega_2,w)a(\omega_3,w) \Bigg\}~~,
    \end{aligned}$}%
    \label{equ:W02-W22}
\end{equation}
while the first line matches with (\ref{equ:w1+infty-4D}), there are extra terms. We will now see that the contribution from cubic operators exactly cancels them. 
From (\ref{equ:W1s-mode}) and (\ref{equ:W23-mode}), one can show that for general $s$ we have
\begin{equation}
    \begin{aligned}
       \Big[W_s^{1,-}(z,\bz), W_{2,matter}^{3,-}&(w,\bw)\Big] ~=~ -\,\frac{3}{2}\frac{1}{4}\frac{1}{(2\pi)^3}\frac{(-i)^s}{s!}\,D_w^{s+2}\delta^{(2)}(z-w)\,\int_0^{\infty}d\omega_1 \int_0^{\infty}d\omega_2\,\omega_2^2\\
       &\Bigg\{ \frac{1}{s+1}\,\pa_{\omega_1}^{s+1}\delta(\omega_1-\omega_2) + \omega_1^{-1}\pa_{\omega_1}^{s}\delta(\omega_1-\omega_2) \Bigg\}\,a^{\dagger}(\omega_1,w,\bw)a(\omega_2,w,\bw)~~.
    \end{aligned}
    \label{equ:qs1-q23}
\end{equation}
We leave the explicit derivation of this result in Appendix~\ref{app:calcs}. When $s=0$, this reduces to
\begin{equation}
    \begin{split}
        \Big[W_0^{1,-}(z,\bz), W_{2,matter}^{3,-}(w,\bw)\Big] ~=&~ -\,\frac{3}{2}\frac{1}{4}\frac{1}{(2\pi)^3}\,D_w^{2}\delta^{(2)}(z-w)\,\int_0^{\infty}d\omega_1 \int_0^{\infty}d\omega_2\,\omega_2^2\\
       &~\Bigg\{ \,\pa_{\omega_1}\delta(\omega_1-\omega_2) + \omega_1^{-1}\delta(\omega_1-\omega_2) \Bigg\}\,a^{\dagger}(\omega_1,w,\bw)a(\omega_2,w,\bw)~~,
    \end{split}
\end{equation}
which cancels the extra terms in (\ref{equ:W02-W22}). Since $W^3_{0,matter}=0$, we can add these together to get
\begin{equation}\resizebox{0.9\textwidth}{!}{$%
    \begin{aligned}
        \Big[ W_0(z,\bz),\,W_{2}(w,\bw) \Big]^2_{matter} ~=&~  \Big[ W^2_{0,matter}(z,\bz),\,W^2_{2,matter}(w,\bw) \Big]  
        ~+~  \Big[ W^1_0(z,\bz),\,W^3_{2,matter}(w,\bw) \Big] \\
        ~=&~ \frac{i}{2}\,\Big[ 4\,\pa_{w}\delta^{(2)}(z-w)\,W^2_{1}(w,\bw) + \delta^{(2)}(z-w)\,\pa_w\,W^2_{1}(w,\bw) \Big]_{matter} ~~,
    \end{aligned}$}%
\end{equation}
as desired.

\paragraph{$\mathbf{\succ}~s=1, s'=2:$}
Direct computation gives
\begin{equation}\resizebox{0.9\textwidth}{!}{$%
    \begin{aligned}
       \Big[W_{1,matter}^{2,-}(z), W_{2,matter}^{2,-}(w)\Big] ~=&~ \frac{i}{2}\,\Big[ 5\,D_{w}\delta^{(2)}(z-w)\,W^{2,-}_{2,matter}(w) + 2\,\delta^{(2)}(z-w)\,D_w\,W^{2,-}_{2,matter}(w) \Big]\\
       &~+~ \frac{3}{2}\frac{1}{4}\frac{-i}{(2\pi)^3}\,D_w^{3}\delta^{(2)}(z-w)\,\Bigg\{ \int_0^{\infty}d\omega\,a^{\dagger}(\omega,w)a(\omega,w)\\
       & ~+~  \int_0^{\infty}d\omega_1\int_0^{\infty}d\omega_2\,\omega_2\,\pa_{\omega_1}\delta(\omega_1-\omega_2)\,a^{\dagger}(\omega_1,w)a(\omega_2,w) \\
       &~+~ \frac{1}{2} \int_0^{\infty}d\omega_1\int_0^{\infty}d\omega_2\,\omega^2_2\,\pa^2_{\omega_1}\delta(\omega_1-\omega_2)\,a^{\dagger}(\omega_1,w)a(\omega_2,w) \Bigg\}~~.
    \end{aligned}$}%
    \label{equ:W12-W22}
\end{equation}
Plugging $s=1$ into (\ref{equ:qs1-q23}) and using
\begin{equation}
    \omega_1^{-1}\pa_{\omega_1}\delta(\omega_1-\omega_2) ~=~ \omega_2^{-1}\pa_{\omega_1}\delta(\omega_1-\omega_2) + \omega_2^{-2}\delta(\omega_1-\omega_2) ~~,
\end{equation}
we can see that $[ W^1_1(z,\bz),\,W^3_{2,matter}(w,\bw)]$ exactly cancels the last three terms in (\ref{equ:W12-W22}). Again, noting that $W^3_{1,matter}=0$, we have
\begin{equation}\resizebox{0.9\textwidth}{!}{$%
    \begin{aligned}
        \Big[ W_1(z,\bz),\,W_{2}(w,\bw) \Big]^2_{matter} ~=&~  \Big[ W^2_{1,matter}(z,\bz),\,W^2_{2,matter}(w,\bw) \Big]   
        ~+~  \Big[ W^1_1(z,\bz),\,W^3_{2,matter}(w,\bw) \Big] \\
        ~=&~ \frac{i}{2}\,\Big[ 5\,D_{w}\delta^{(2)}(z-w)\,W^2_{2}(w,\bw) + 2\,\delta^{(2)}(z-w)\,D_w\,W^2_{2}(w,\bw) \Big]_{matter} ~~,
    \end{aligned}$}%
\end{equation}
matching (\ref{equ:w1+infty-4D}). 

\vspace{1em}

In sum, we have demonstrated closure of the matter algebra up to quadratic order, checked that it reproduces the correct graviton-matter OPE and, in doing so, identified an extended class of detector/light-ray operators corresponding to the $w_{1+\infty}$ symmetry of asymptotically flat gravity.

 \section{Discussion}\label{sec:discussion}
Here we've merged results from the conformal collider and celestial holography literature to, first, recast various 4D light-ray operators that have been used extensively elsewhere in a manifestly celestial language; and, second, extend these collider constructions based on recent celestial results. Concretely, we saw that we can rephrase the ANEC and other related light-ray operators as falling within the conformally soft sector of the matter CFT. This becomes clear when we view CCFT from an extrapolate dictionary perspective, where for massless excitations we are organizing correlation functions of operators on null infinity in terms of boost eigenstates.

The known examples of these matter-sector hard charges are then seen to fill the first entries of a tower of $\Delta=3$ modes. Befittingly, tools developed for celestial CFT from a third, corner symmetry, approach proved key to demonstrating explicitly that we can extend this tower to complete the matter contributions to the generators of the newfound $Lw_{1+\infty}$ symmetry of gravity. We showed how the quadratic modes reproduce the correct universal coupling of conformally soft gravitons to matter fields, previously identified from collinear limits of scattering. These modes are a natural generalization of both the ${\cal L}_n$ and ${\cal E}_J$ detector operators in the conformal collider literature. In the course of these derivations, it is hard not to appreciate the power of these symmetries to pinpoint the relevant series of detector-like operators. Demanding that this algebra closes through quadratic order further highlights a distinction between the remainder of the tower and its BMS subalgebra. While the matter and gravity representations decouple into mutually commuting algebras, the remainder does not. This highlights the importance of having a massless spin-2 mode in the bulk in order to realize these extra symmetries. 

We hope that this initial foray into celestial conformal colliders, via the concrete example of showing that one can realize the $w_{1+\infty}$ symmetry in the matter sector coupled to gravity, is just the beginning of a rich interplay between these programs.

\section*{Acknowledgements}
We would like to thank Luca Ciambelli, Laurent Freidel, and Justin Kulp for useful conversations. The research of YH and SP is supported by the Celestial Holography Initiative at the Perimeter Institute for Theoretical Physics. Research at the Perimeter Institute is supported by the Government of Canada through the Department of Innovation, Science and Industry Canada and by the Province of Ontario through the Ministry of Colleges and Universities.

\appendix
\section{4D Conformal Generators in Various Coordinates}\label{app:4Dcoords}
In this appendix, we will show explicitly how the 4D conformal generators in different coordinate systems relate to one another. First, the conformal algebra reads
\begin{align}
    &[P_{\mu},P_{\nu}] ~=~ [K_{\mu},K_{\nu}] ~=~ [D,D] ~=~[D,M_{\mu\nu}] ~=~  0~~, \\
    &[P_{\mu},D] ~=~ -i\,P_{\mu} ~~,~~ [K_{\mu},D] ~=~ i\,K_{\mu} ~~,~~ [P_{\mu},K_{\nu}] ~=~ -2i\,(\eta_{\mu\nu}\,D + M_{\mu\nu}) ~~,\\
    &[P_{\mu},M_{\rho\sigma}] ~=~ i\,\eta_{\mu[\rho}\,P_{\sigma]} ~~,~~[K_{\mu},M_{\rho\sigma}] ~=~ i\,\eta_{\mu[\rho}\,K_{\sigma]} ~~,~~\\
    &[M_{\mu\nu},M^{\rho\sigma}] ~=~ i\,\delta_{[\mu}{}^{[\rho}\,M^{\sigma]}{}_{\nu]}~~.
\end{align}
We will now discuss the vector field representation of this algebra in various coordinate bases.

\paragraph{Cartesian Coordinates $\{x^0,x^1,x^2,x^3\}$}
\begin{align}
    P_{\mu} ~=&~ -i\,\pa_{\mu} ~~,\\
    M_{\mu\nu} ~=&~ i\,(x_{[\mu}\,\pa_{\nu]})~~,\\
    D ~=&~ -i\,x^{\mu}\,\pa_{\mu}~~,\\
    K_{\mu} ~=&~ i\,(x^2\,\pa_{\mu} - 2\,x_{\mu}\,x_{\nu}\,\pa^{\nu})~~,
\end{align}
where $\mu = 0,1,2,3$ and the metric is
\begin{equation}
    \eta_{\mu\nu} ~=~ \begin{pmatrix}
    -1 & 0 & 0 & 0\\
    0 & 1 & 0 & 0\\
    0 & 0 & 1 & 0\\
    0 & 0 & 0 & 1
    \end{pmatrix}~~.
\end{equation}

\paragraph{Light-cone Coordinates $\{x^+,x^-,x^1,x^2\}$}
The light-cone coordinate is defined as follows
\begin{equation}
    x^+ ~=~ (x^0+x^3) ~~,~~ x^- ~=~ (x^0-x^3) ~~,~~ x^1 ~=~ x^1 ~~,~~ x^2 ~=~ x^2~~,
\end{equation}
and, inversely,
\begin{equation}
    x^0 ~=~ \frac{1}{2}\,(x^++x^-) ~~,~~ x^3 ~=~ \frac{1}{2}\,(x^+-x^-)~~.
\end{equation}
We will use $a=+,-,1,2$ to label the light-cone coordinates. Namely,  
\begin{align}
    D ~=&~ -i\,x^{\mu}\,\pa_{\mu} ~=~ -i\,x^{a}\,\pa_{a} ~~,\\
    M^{03} ~=&~ i\,(x^0\pa^3-x^3\pa^0) ~=~ i\,(x^0\pa_3+x^3\pa_0) ~=~ i\,(x^+\pa_+-x^-\pa_-) ~~,\\
    M^{12} ~=&~ i\,(x^1\pa^2-x^2\pa^1) ~=~ i\,(x^1\pa_2-x^2\pa_1)  ~~.
\end{align}

\paragraph{`Inverse' light-cone coordinates $\{y^+,y^-,y^1,y^2\}$} Following \cite{Hofman:2008ar}, the `inverse' light-cone coordinates are defined via the map
\begin{equation}
    y^+ ~=~ -\frac{1}{x^+} ~~,~~
    y^- ~=~ x^- - \frac{x_1^2+x_2^2}{x^+} ~~,~~
    y^1 ~=~ \frac{x^1}{x^+} ~~,~~
    y^2 ~=~ \frac{x^2}{x^+}~~,
\end{equation}
and, inversely,
\begin{equation}
    x^+ ~=~ -\frac{1}{y^+} ~~,~~
    x^- ~=~ y^- - \frac{y_1^2+y_2^2}{y^+} ~~,~~
    x^1 ~=~ -\,\frac{y^1}{y^+} ~~,~~
    x^2 ~=~ -\,\frac{y^2}{y^+}~~.
\end{equation}
We can explicitly see that the dilatation $D$ in the $x^a$ coordinates is the boost $M^{03}$ in $y^a$ coordinates
\begin{equation}
    \begin{split}
        D(x^a) ~=&~ -i\,x^{a}\,\pa_{a} ~=~ -i\,x^{+}\,\pa_{x^+} -i\,x^{-}\,\pa_{x^-} -i\,x^{1}\,\pa_{x^1} -i\,x^{2}\,\pa_{x^2}\\
        ~=&~ -i\,x^{+}\,\left(\pa_{y^+}\frac{\pa y^+}{\pa x^+} + \pa_{y^-}\frac{\pa y^-}{\pa x^+} + \pa_{y^1}\frac{\pa y^1}{\pa x^+} + \pa_{y^2}\frac{\pa y^2}{\pa x^+}\right) -i\,x^{-}\,\left( \pa_{y^-}\frac{\pa y^-}{\pa x^-} \right) \\
        &~ -i\,x^{1}\,\left(\pa_{y^-}\frac{\pa y^-}{\pa x^1} + \pa_{y^1}\frac{\pa y^1}{\pa x^1} \right) -i\,x^{2}\,\left(\pa_{y^-}\frac{\pa y^-}{\pa x^2} + \pa_{y^2}\frac{\pa y^2}{\pa x^2} \right)\\
        ~=&~ i\,(y^+\,\pa_{y^+} - y^-\,\pa_{y^-}) ~=~ M^{03}(y^a) ~~,
    \end{split}
\end{equation}
and 
\begin{equation}
    \begin{split}
        M^{03}(x^a) ~=&~ i\,(x^+\pa_{x^+}-x^-\pa_{x^-})\\
        ~=&~ i\,x^{+}\,\left(\pa_{y^+}\frac{\pa y^+}{\pa x^+} + \pa_{y^-}\frac{\pa y^-}{\pa x^+} + \pa_{y^1}\frac{\pa y^1}{\pa x^+} + \pa_{y^2}\frac{\pa y^2}{\pa x^+}\right) -i\,x^{-}\,\left( \pa_{y^-}\frac{\pa y^-}{\pa x^-} \right) \\
        ~=&~ -i\,(y^{+}\,\pa_{y^+} + y^{-}\,\pa_{y^-} + y^{1}\,\pa_{y^1} + y^{2}\,\pa_{y^2}) ~=~ D(y^a)~~.
    \end{split}
\end{equation}
Since we define the boost weight in the $x^+,x^-$ plane as the 2D conformal weight, we see that the celestial conformal weight is equal to the 4D conformal weight in the $y^a$ basis. 
\begin{equation}
    M^{03}(x^a) ~=~ \Delta_{\rm 2D} ~=~ \Delta_{\rm 4D}(y^a) ~~.
\end{equation}
Furthermore, the spin in the transverse  $y^1,y^2$ plane is 2D spin. We see that
\begin{equation}
\begin{split}
     M^{12}(y^a) ~=&~ i\,(y^1\pa_{y^2}-y^2\pa_{y^1}) \\
     ~=&~ i\, y^{1}\,\left(\pa_{x^-}\frac{\pa x^-}{\pa y^2} + \pa_{x^2}\frac{\pa x^2}{\pa y^2} \right) - i\,y^{2}\,\left(\pa_{x^-}\frac{\pa x^-}{\pa y^1} + \pa_{x^1}\frac{\pa x^1}{\pa y^1} \right)\\
     ~=&~ i\,(x^1\pa_{x^2}-x^2\pa_{x^1})  ~=~  M^{12}(x^a) ~~,
\end{split}
\end{equation}
therefore the $J^{12}$ spin is the same in both $x^a$ and $y^a$ coordinate systems.

\section{Useful Identities}\label{app:identities}
In this appendix, we list several identities that are useful in the derivations throughout this paper.

\paragraph{Fourier transform}
\begin{align}
    \int_{-\infty}^{+\infty}\,du\,e^{i(\omega_1-\omega_2)u} ~=&~ 2\pi\,\delta(\omega_1-\omega_2)~~,
    \label{equ:exp-Fourier} \\
    \int_{-\infty}^{+\infty}\,du\,u^{m}\,e^{i(\omega-\omega')u} ~=&~ (2\pi)\,(-i\pa_{\omega})^{m}\,\delta(\omega-\omega') ~~.
    \label{equ:deriv-exp-Fourier}
\end{align}

\paragraph{Delta function identities}
\begin{align}
    \pa_z\delta^{(2)}(z-w)~=&~ -\pa_w\delta^{(2)}(z-w)~~, \label{equ:deltafun-deriv}\\
     \pa_x^n\,\delta(x) ~=&~ \frac{(-1)^n\,n!}{x^n}\,\delta(x) ~~, \label{equ:deriv-deltafunc} \\
    \int dx\, \delta'(x)\,f(x) ~=&~ -\,\int dx\,\delta(x)\,f'(x)~~.
    \label{equ:deltafunc-deriv-integral}
\end{align}
Meanwhile, (\ref{equ:deriv-deltafunc}) further yields the following recursion relation
\begin{equation}
    \frac{\pa_x^{n+1}\,\delta(x)}{n+1} ~=~ -\,\frac{\pa_x^n\,\delta(x)}{x} ~~.
    \label{equ:deltafunc-deriv-recursion}
\end{equation}
Another helpful manipulation that we have used many times throughout this paper is
\begin{equation}
\begin{split}
    \pa_z\delta^{(2)}(z-w)\,f(w,\bw) ~=&~ \pa_z\,\Big[\delta^{(2)}(z-w)\,f(w,\bw)\Big] ~=~ \pa_z\,\Big[\delta^{(2)}(z-w)\,f(z,\bz)\Big] \\
    ~=&~ \pa_z\delta^{(2)}(z-w)\,f(z,\bz) ~+~ \delta^{(2)}(z-w)\,\pa_zf(z,\bz)~~.
\end{split}
    \label{equ:delfuc-manipulation}
\end{equation}

\paragraph{Gamma function identities}~\cite{Freidel:2021ytz}
\begin{equation}
    \Gamma(\a-n) ~=~ (-1)^{n-1}\,\frac{\Gamma(-\a)\Gamma(1+\a)}{\Gamma(n+1-\a)} ~~,n\in\mathbb{Z}~.
    \label{equ:GammafuncID}
\end{equation}

\paragraph{Integral identities}
\begin{equation}
    \int^{+\infty}_{0}\,du\,e^{i\omega u} ~=~  \frac{i}{\omega} ~~,
    \label{equ:half-exp-Fourier}
\end{equation}

\begin{equation}
\int_0^{\infty}\,d\omega\,\omega^m\,\pa_{\omega}^k\delta(\omega-\omega_p)\,\pa_{\omega_q}^l\delta(\omega_q-\omega) ~=~ \sum_{r=0}^k\,(-1)^r\,\begin{pmatrix}
k\\
r
\end{pmatrix}\,(m)_r\,\omega_p^{m-r}\,\pa_{\omega_q}^{k+l-r}\delta(\omega_q-\omega_p)~,
    \label{equ:two-deriv-delta-integralID}
\end{equation}
where $(m)_r=m(m-1)\cdots(m-r+1)$ is the falling factorial. This identity can be derived by using (\ref{equ:deltafun-deriv}) and (\ref{equ:deltafunc-deriv-integral}) repeatedly. Equivalently, if we refer to the original integral as $A(m,k,l)$, applying (\ref{equ:deltafun-deriv}) and (\ref{equ:deltafunc-deriv-integral}) once yields the following recursion relation
\begin{equation}
    A(m,k,l) ~=~ (-m)\,A(m-1,k-1,l) ~+~ A(m,k-1,l+1)~~,
\end{equation}
which has the solution
\begin{equation}
    A(m,k,l) ~=~ \sum_{r=0}^k\,(-1)^r\,\begin{pmatrix}
k\\
r
\end{pmatrix}\,(m)_r\,A(m-r,0,k+l-r)~~.
\end{equation}
Note that 
\begin{equation}
    \begin{split}
        A(m-r,0,k+l-r) ~=&~ \int_0^{\infty}\,d\omega\,\omega^{m-r}\,\delta(\omega-\omega_p)\,\pa_{\omega_q}^{k+l-r}\delta(\omega_q-\omega)\\
        ~=&~ \omega_p^{m-r}\,\pa_{\omega_q}^{k+l-r}\delta(\omega_q-\omega) ~~,\\
    \end{split}
\end{equation}
and we arrive at (\ref{equ:two-deriv-delta-integralID}).

\section{\texorpdfstring{Derivation of $[W_s^1, W_{2,matter}^3]$}{Derivation of [Ws, W2]}}\label{app:calcs}

In this appendix, we include additional steps for the derivation of~\eqref{equ:qs1-q23}, so that the reader may more readily follow along. First, recall that 
\begin{equation}\scalemath{.97}{
  \badat{3}
         W_s^{1,-}(z,\bz) ~=&~ \frac{1}{2}\frac{(-1)^{s+2}}{s!}\,\frac{i^s}{(2\pi)}\,\int_0^{\infty}d\omega\,\omega\,\pa_{\omega}^s\delta(\omega)\,D_z^{s+2}\,a_-(\omega,z,\bz)~,   
         \eadat}
\end{equation}
and
\begin{equation}\scalemath{.97}{
   \badat{3}
        W_{2,matter}^{3,-}(z,\bz) 
   ~=&~ \frac{3}{2}\frac{1}{4}\,\frac{i}{(2\pi)^6}\,\int_{-\infty}^{+\infty}\,du\,\int_0^{\infty}d\omega_1\,a_-^{\dagger}(\omega_1,z,\bz)\,e^{i\omega_1u}\int_{+\infty}^{u}\,du'\\
       &\qquad\qquad \int_0^{\infty}d\omega_2\,a^{\dagger}(\omega_2,z,\bz)\, \int_0^{\infty}d\omega_3\,\omega_3^2\,e^{i(\omega_2-\omega_3)u'}\,a(\omega_3,z,\bz)\\
       &~+~ \frac{3}{2}\frac{1}{4}\,\frac{i}{(2\pi)^6}\,\int_{-\infty}^{+\infty}\,du\,\int_0^{\infty}d\omega_1\,a^{\dagger}(\omega_1,z,\bz)\,e^{i\omega_1u}\int_{+\infty}^{u}\,du'\\
       &\qquad\qquad \int_0^{\infty}d\omega_2\,a_-^{\dagger}(\omega_2,z,\bz)\, \int_0^{\infty}d\omega_3\,\omega_3^2\,e^{i(\omega_2-\omega_3)u'}\,a(\omega_3,z,\bz)~.\\
    \eadat}
\end{equation}
Since $a_-$ commutes with everything except for $a_-^{\dagger}$, we can use our canonical commutation relations~\eqref{eq:adaggera} to evaluate the commutator 
\begin{equation}\scalemath{.93}{
    \badat{3}
         &\Big[W_s^{1,-}(z,\bz), W_{2,matter}^{3,-}(w,\bw)\Big] 
        ~=~
        \frac{(-1)^{s+2}}{s!}\,\frac{3}{2}\frac{1}{4}\,\frac{i^{s+1}}{(2\pi)^4}D_z^{s+2}\delta^{(2)}(z-w)\,\int_{-\infty}^{+\infty}\,du\int_0^{\infty}d\omega_2\,e^{i\omega_2 u}\,\pa_{\omega_2}^s\delta(\omega_2)\\
        &~~~~~~\qquad\qquad\qquad \int_0^{\infty}d\omega_3 \int_0^{\infty}d\omega_4\,\omega_4^2\,\int_{+\infty}^{u}\,du'\,e^{i(\omega_3-\omega_4)u'}\,a^{\dagger}(\omega_3,w,\bw)a(\omega_4,w,\bw)\\
        &~~~~~~\qquad\qquad\qquad+\frac{(-1)^{s+2}}{s!}\,\frac{3}{2}\frac{1}{4}\,\frac{i^{s+1}}{(2\pi)^4}D_z^{s+2}\delta^{(2)}(z-w)\,\int_{-\infty}^{+\infty}\,du\,\int_0^{\infty}d\omega_2\,e^{i\omega_2u}\,\int_0^{\infty}d\omega_3\,\pa_{\omega_3}^s\delta(\omega_3)\\
        &~~~~~~\qquad\qquad\qquad \int_0^{\infty}d\omega_4\,\omega_4^2\,\int_{+\infty}^{u}\,du'\, e^{i(\omega_3-\omega_4)u'}\,a^{\dagger}(\omega_2,w,\bw)a(\omega_4,w,\bw)~.
    \eadat}
    \label{equ:C.34}
\end{equation}
Now let's look at these two integrals, in turn. First, we have
\begin{equation}
\scalemath{.93}{
    \badat{3}
        & \int_{-\infty}^{+\infty}\,du\int_0^{\infty}d\omega_2\,e^{i\omega_2 u}\,\pa_{\omega_2}^s\delta(\omega_2)\,\int_{+\infty}^{u}\,du'\,e^{i(\omega_3-\omega_4)u'} \\
        ~&\qquad\qquad\qquad=~ -\,\int_{-\infty}^{+\infty}\,du\int_0^{\infty}d\omega_2\,e^{i\omega_2 u}\,\pa_{\omega_2}^s\delta(\omega_2)\,\int^{+\infty}_{u}\,du'\,e^{i(\omega_3-\omega_4)u'}\\
        ~&\qquad\qquad\qquad=~ -\,\int_{-\infty}^{+\infty}\,du\int_0^{\infty}d\omega_2\,\,e^{i(\omega_2+\omega_3-\omega_4)u}\,\pa_{\omega_2}^s\delta(\omega_2)\,\int^{+\infty}_{0}\,du'\,e^{i(\omega_3-\omega_4)u'} \\
        ~&\qquad\qquad\qquad=~ -\,(2\pi)\,\int_0^{\infty}d\omega_2\,\pa_{\omega_2}^s\delta(\omega_2)\,\delta(\omega_2+\omega_3-\omega_4)\,\frac{i}{\omega_3-\omega_4} \\
        ~&\qquad\qquad\qquad=~ (-1)^s\,(2\pi)\,\frac{-i}{\omega_3-\omega_4}\,\pa_{\omega_3}^s\delta(\omega_3-\omega_4) ~=~ \frac{i}{s+1}\,(-1)^s\,(2\pi)\,\pa_{\omega_3}^{s+1}\delta(\omega_3-\omega_4)~,
    \eadat}
\end{equation}
where we have used the identities (\ref{equ:exp-Fourier}), (\ref{equ:deltafunc-deriv-recursion}), and (\ref{equ:half-exp-Fourier}). 
Similarly, for the second integral we can do the following manipulation 
\begin{equation}
    \begin{split}
        & \int_{-\infty}^{+\infty}\,du\,e^{i\omega_2u}\,\int_0^{\infty}d\omega_3\,\pa_{\omega_3}^s\delta(\omega_3)\,\int_{+\infty}^{u}\,du'\, e^{i(\omega_3-\omega_4)u'} \\
        ~&\qquad\qquad\qquad=~ -\,\int_{-\infty}^{+\infty}\,du\,\int_0^{\infty}d\omega_3\,\pa_{\omega_3}^s\delta(\omega_3)\,e^{i(\omega_2+\omega_3-\omega_4)u}\,\int^{+\infty}_{0}\,du'\, e^{i(\omega_3-\omega_4)u'} \\
        ~&\qquad\qquad\qquad=~ -\,(2\pi)\,\int_0^{\infty}d\omega_3\,\pa_{\omega_3}^s\delta(\omega_3)\,\delta(\omega_2+\omega_3-\omega_4)\,\frac{i}{\omega_3-\omega_4}\\
        ~&\qquad\qquad\qquad=~ (-1)^s\,(2\pi)\,i\,\omega_2^{-1}\,\pa_{\omega_2}^{s}\delta(\omega_2-\omega_4)~.
    \end{split}
\end{equation}
Finally, substituting these two integrals back into (\ref{equ:C.34}), and using the identity 
\begin{equation}
    D_z^{s+2}\delta^{(2)}(z-w)~=~(-1)^{s+2}\,D_w^{s+2}\delta^{(2)}(z-w)~,
\end{equation}
we find that
\begin{equation}
    \begin{aligned}
       \Big[W_s^{1,-}(z,\bz), W_{2,matter}^{3,-}&(w,\bw)\Big] ~=~ -\,\frac{3}{2}\frac{1}{4}\frac{1}{(2\pi)^3}\frac{(-i)^s}{s!}\,D_w^{s+2}\delta^{(2)}(z-w)\,\int_0^{\infty}d\omega_1 \int_0^{\infty}d\omega_2\,\omega_2^2\\
       &\Bigg\{ \frac{1}{s+1}\,\pa_{\omega_1}^{s+1}\delta(\omega_1-\omega_2) + \omega_1^{-1}\pa_{\omega_1}^{s}\delta(\omega_1-\omega_2) \Bigg\}\,a^{\dagger}(\omega_1,w,\bw)a(\omega_2,w,\bw)~.
    \end{aligned}
\end{equation}

\bibliographystyle{utphys}
\bibliography{references}

\end{document}